\documentclass[%
preprint,
 aps,
]{revtex4-1}

\usepackage{graphicx}% Include figure files
\usepackage{dcolumn}% Align table columns on decimal point
\usepackage{bm}% bold math
\usepackage{amssymb,amsmath,amsthm,accents}
\usepackage{amscd,amsfonts}
\usepackage{fullpage}
\usepackage{float}
\usepackage{lineno}
\usepackage{mathrsfs}
\usepackage{hyperref}
\usepackage{url}
\usepackage{color}
\usepackage{xcolor}
\usepackage{epsfig}
\usepackage{epstopdf}
\DeclareGraphicsExtensions{.eps,.pdf,.png,.jpg,.jpeg}
\usepackage{feynmp}

\usepackage{float}
\usepackage{graphicx}
\usepackage{subcaption}
\usepackage{color}
\usepackage{rotating}
\usepackage{multirow}
\usepackage{afterpage}
\usepackage{footnote}
\usepackage[T1]{fontenc}
\usepackage{ae,aecompl}
\begin{document}

\title{On the flat galactic rotational curves in $\mathnormal{f(R)}$ gravity
	%Dark matter in $f(\mathcal{R})$ gravity
}% Force line breaks with \\
	\author{Muhammad Usman}
	\email{muhammad\_usman\_sharif@yahoo.com}
	\email{m.usman@comsats.edu.pk}
	\affiliation{
		Department of Physics,
		COMSATS University Islamabad (CUI), \\ Islamabad Campus,
		Park Road, Tarlai Kalan, \\ Islamabad 45550, Pakistan.
	}
	
	\date{\today}
	% % % % % % % % % % % % % % % % % % % % % % % % % % % % % % % % % % % % % % % % % % % % % % % % % % % % % % % % % % % % % % % % % % % % % % % % % % % % % % % % % % % % % %
	\begin{abstract}
	A mysterious dark matter is supposed to exist in the galactic halos. In this contrast, we discuss the possibility of explaining the flat rotational velocity curves in f(R) gravity by solving field equations numerically in vacuum and for different matter distributions. For a spherically symmetric static space-time (as the galactic environment) we give metric for constant rotational velocity regions. For a constant rotational velocity region, we prove that all values of rotational velocities (most importantly observed rotational velocity ~200-300Km/s) do not lead to an analytic solution of the vacuum field equations. We then obtain numerical solutions of the field equations in vacuum and for three types of mass distributions named: (1) power law density profile, (2) simple model for galaxy with a core and, (3) Navarro, Frank and White (NFW) profile, for M31 and Milky way galaxy. The solutions suggest a slight modification from linear relations from R for vacuum whereas a significant deviation from R for the distributions can give flat rotational curves. Using Brans-Dicke theory, we also relate obtained modified gravity function with the equivalent scalar fields, the procedure gives us very interesting phenomena and behavior of dark matter in the galactic environment. We observe that the scalar dark matter, coming from different modified gravity functions of matter profiles, does not accumulate as the baryonic matter. These results then can be used to explain the spatial offset of the center of the total mass from the center of the baryonic mass peaks of the bullet cluster and Abell-520.
	\end{abstract}
	% % % % % % % % % % % % % % % % % % % % % % % % % % % % % % % % % % % % % % % % % % % % % % % % % % % % % % % % % % % % % % % % % % % % % % % % % % % % % % % % % % % % % %
	\maketitle
	% % % % % % % % % % % % % % % % % % % % % % % % % % % % % % % % % % % % % % % % % % % % % % % % % % % % % % % % % % % % % % % % % % % % % % % % % % % % % % % % % % % % % %
	% % % % % % % % % % % % % % % % % % % % % % % % % % % % % % % % % % % % % % % % % % % % % % % % % % % % % % % % % % % % % % % % % % % % % % % % % % % % % % % % % % % % % %
	
	\section{Introduction}
	There is no priori reason to consider that the gravity should be linear in Ricci scalar $\mathcal{R}$ except it best matches with the %solar system 
	observations (excluding dark matter and dark energy). Dark energy is related to the mystery of accelerated expanding Universe whereas dark matter is related to the observed rotation velocity curves of the galaxy. Here, we address the problem of the dark matter. The Newtonian dynamics and General Relativity (GR) can not explain these phenomena which leads to idea that either one or both sides of the Einstein field equations (EFE) are incomplete. The right hand side of the EFE comprise of the matter part whereas left side is usually taken to be the geometric part. According to John A. Wheeler EFEs tell ``Spacetime tells matter how to move; matter tells spacetime how to curve'' \citet{wheeler2010geons}. There are many proposals that address the dark matter problem. Few of them include weakly interacting massive particles (WIMP) \citet{Jungman1996195,PhysRevD.76.043520}, scalar field from extensions of standard model of Particle Physics (SM) \citet{PhysRevD.82.123533}, modified Newtonian dynamics (MOND) \citet{Milgrom:1983ca} and modified gravity \citet{Bohmer2008386,refId0-1,refId0-2,Jamil2012}. In modified gravity, one alters the left hand side (geometric part) of the EFE. It then can be used to explain the observed rotational velocity curves to be a geometric effect only.
	
	The dark matter is the type of matter supposed to be surrounding galaxies that gives the observed rotational velocity curves. Einstein-Hilbert action could also be one of the reasons that there is some discrepancy between observed and expected tangential velocity of a particle moving around a galactic center i.e. the gravity is not linear at the galactic scale \footnote{Since the GR is a linear theory of gravity and it cannot accord for the observed rotational curves therefore a gravitational theory which can explain observed rotational curves must be nonlinear.}.
	
	Here, we present a $f(\mathcal{R})$ gravity model by solving field equations numerically that can describe the dark matter as a geometric effect. The scheme of the article is as follows; using a metric that can describe the constant velocity region of space around a galaxy in vacuum we solve the field equations for the metric in $f(\mathcal{R})$ gravity numerically and show that analytical solution for all possible values of tangential velocities is not possible. We know that there is some observable (normal) matter around the galaxy by using which we could actually get some value of the tangential velocity of a particle around the galaxy. If we get some $f(\mathcal{R})$ function for constant velocity region around the galaxy in vacuum, then the difference in the actual $f(\mathcal{R})$ function and general relativity that could replace the dark matter must have to be less than the difference between the obtained $f(\mathcal{R})$ function and general relativity $\big(~$i.e. $|\mathnormal{f}(\mathcal{R})_{\text{actual}}-\mathcal{R}|<|\mathnormal{f}(\mathcal{R})_{\text{obtained}}-\mathcal{R}|$$~\big)$ because of some contribution from normal matter\footnote{The rotational velocity curves of galaxy is due to both normal as well as dark matter. Thus solving the EFEs in vacuum would give us the upper bound on the amount of dark matter.}. 
	
	Previously while solving the field equations in vacuum, we found in \citet{Usman2016} that we arrive at two different sets of metric parameters which then give two different forms of $f(\mathcal{R})$ function that could explain the constant tangential velocity curves. The forms indicate that rather than having a positive correction of the form $\mathcal{R}^{1+\alpha}$, a negative correction of the form $\mathcal{R}^{1-\alpha}$ is more likely to happen. 
	
	In this article, we show that in certain situations (for particular tangential velocities domain), it is not possible to obtain an analytic solution of the vacuum field equations in $f(\mathcal{R})$ gravity.
	We then solve the field equations numerically in vacuum and for different matter distributions. This is followed by obtaining the dark matter scalar field and its potential by using Brans-Dicke theory.
	The results suggest a slight deviation from Einstein's gravity for vacuum and significant for matter distributions, could explain the discrepancy between observed and expected rotational velocity curves. 
	
	The article is organized as follows; in the section II, we give a metric that can describe the constant velocity region of space around a galaxy %in vacuum 
	and the field equations for the metric in $f(\mathcal{R})$ gravity. In section III, we give the proofs that analytical solution for all possible values of tangential velocities is not possible. We then present the numerical solutions of the field equations in section IV. The solution of the field equations is followed by the derivation of the scalar field and its potential inspired from the obtained results using Brans-Dicke theory in section V. Section VI is the conclusion and discussion.
	%%%%%%%%%%%%%%%%%%%%%%%%%%%%%%%%%%%%%%%%%%%%%%%%%%%%%%%%%%%%%%%%%%%%%%%%%%%%%%%%%%%%%%%%%
	\section{$f(\mathcal{R})$ gravity field equations}\label{fRfieldequations}
	
	The metric which describes the static and spherically symmetry in the constant rotational velocity regions is
	\begin{equation}\begin{split}\label{eq:metric}
	{\mathrm{d}s}^2=-\mathnormal{e}^{\nu(r)}{\mathrm{d}t}^2%&
	+\mathnormal{e}^{\lambda(r)}{\mathrm{d}r}^2%\\&
	+r^2{\mathrm{d}\theta}^2+r^2{\sin^2\theta}{\mathrm{d}\phi}^2.
	\end{split}\end{equation}
	The metric coefficient $\nu(r)$ can be calculated from the Euler-Lagrange equations of the Lagrangian of the metric. From Euler-Lagrange equations, we get \citet{Bohmer2008386}
	\begin{equation}\label{eq:tangential-velocity}
	\nu'(r)=\dfrac{2}{r}\left(\dfrac{V_t(r)}{c}\right)^2~,
	\end{equation}
	$V_{t}$ being the rotational velocity of a particle around galaxy and $c$ is the speed of light.
	The $f(\mathcal{R})$ modified gravity action is written as
	\begin{equation}\label{eq:action}
	S=\dfrac{1}{2\kappa}\int \sqrt{-g} ~ \Big(f(\mathcal{R})-\mathcal{L}_{m}\Big) ~ \mathrm{d}^4x~,
	\end{equation}
	here $f(\mathcal{R})$ is an arbitrary analytic function of Ricci scalar $\mathcal{R}$. The variation of the above action with respect to the metric $g_{\mu\nu}$ gives from \citet{doi:10.1142/S0218271803004407}
	\begin{equation}\begin{split}\label{eq:fieldequations}
	F(\mathcal{R})\mathcal{R}_{\mu\nu}&
	-\dfrac{1}{2}f(\mathcal{R})\mathnormal{g}_{\mu\nu}%\\&
	-\left( \nabla_\mu\nabla_\nu-\mathnormal{g}_{\mu\nu}\square \right)F(\mathnormal{R})=\kappa ~ T_{\mu\nu}~,
	\end{split}\end{equation}
	where $F(\mathcal{R})=\mathrm{d}f/\mathrm{d}\mathcal{R}$. The contraction (or the trace) of the last equation gives
	\begin{equation}\label{eq:fieldequationscontraction}
	F(\mathcal{R})\mathcal{R}-2f(\mathcal{R})+3\square F(\mathnormal{R})=\kappa ~ g^{\mu\nu} T_{\mu\nu}~.
	\end{equation}
	Upon using Eq. (\ref{eq:fieldequationscontraction}) in Eq. (\ref{eq:fieldequations}) we get modified field equations as \citet{PhysRevD.74.064022}
	\begin{equation}\label{eq:modifiedfieldequations}
	F(\mathcal{R})\mathcal{R}_{\mu\nu}-\nabla_\mu\nabla_\nu F - \dfrac{3}{4} \kappa ~ T_{\mu\nu}=\dfrac{1}{4}\mathnormal{g}_{\mu\nu}\left( F\mathcal{R}-\square F \right) ,
	\end{equation}
	moreover differentiation of Eq. (\ref{eq:fieldequationscontraction}) with respect to `$r$' (denoted by {\large{${}{'}$}}) yields
	\begin{equation}
	\mathcal{R}F^{'}-\mathcal{R}^{'}F+3\left(\square F\right)^{'}=\kappa ~ (g^{\mu\nu}T_{\mu\nu})^{'}~.
	\end{equation}
	%Any solution of modified field (\ref{eq:fieldequations}) equations must also satisfy Eq. (\ref{eq:fieldequationscontraction}) since it has been derived from the same equation. 
	From Eq. (\ref{eq:modifiedfieldequations}), we see that when $\mu=\nu$ then $A_\mu \equiv (F\mathcal{R}_{\mu\mu}-\nabla_\mu\nabla_\mu-3/4\text{\space\space} \kappa \text{\space}T_{\mu\mu})/\mathnormal{g}_{\mu\mu}$ implies that $A_\mu-A_\nu=0$ which implies $A_0-A_1=0,~ A_1-A_2=0\text{ and }A_0-A_2=0$. Assuming the perfect fluid with $T_{\mu\nu}=\text{diag}\big(\rho,p,p,p\big)$, where $\rho$ is energy density while $p$ is the pressure of galactic matter (since it is non-relativistic $p=0$). %Since the field equations also give three equation we can use any three equations from all six equations.
	The equations $A_0-A_1=0,~ A_1-A_2=0$, the $rr$ component of the field equations and Eq. (\ref{eq:fieldequationscontraction}) can be written as \citet{refId0-1,refId0-2},
	\begin{equation}\label{eq:1st}
	F^{''}-\dfrac{1}{2}(\nu^{'}+\lambda^{'})F^{'}+\dfrac{1}{r}(\nu^{'}+\lambda^{'})F+\dfrac{3}{4}\kappa ~\text{e}^{-\nu}\rho=0~,
	\end{equation}
	\vspace{-2em}
	\begin{equation}%\left.
	\begin{split}\label{eq:2nd}
	\nu^{''}+{\nu^{'}}^2-\dfrac{1}{2}&\left(\nu^{'}+\dfrac{2}{r}\right)
	\left(\nu^{'}+\lambda^{'}\right)%\\&
	-\dfrac{2}{r^2}+2\dfrac{\mathnormal{e}^{\lambda}}{r^2}=2\dfrac{F^{''}}{F}-\left(\lambda^{'}+\dfrac{2}{r}\right)\dfrac{F^{'}}{F}~,
	\end{split}%\right\}
	\end{equation}
	\vspace{-1.0em}
	\begin{equation}\label{eq:3rd}
	f=F\mathnormal{e}^{-\lambda}\left[\nu^{''}-\dfrac{1}{2}\nu^{'}\left(\nu^{'}+\lambda^{'}\right)- 2\dfrac{\lambda^{'}}{r}%\right. %\\
	+%\left.
	\left(\nu^{'}+\dfrac{4}{r}\right)\dfrac{F^{'}}{F} \right],
	\end{equation}
	\vspace{-1.0em}
	\begin{equation}\begin{split}\label{eq:4th}
	\mathcal{R}=2\dfrac{f}{F} -\dfrac{\kappa}{F} ~ \text{e}^{-\nu}\rho -3& 
	\mathnormal{e}^{-\lambda}\left[\dfrac{F^{''}}{F}\right.%\\&
	+\left( \dfrac{1}{2}\left(\nu^{'}-\lambda^{'}\right)+\dfrac{2}{r} \right)\left.\dfrac{F^{'}}{F} \right],
	\end{split}\end{equation}
	respectively.
	
	Introducing  $\eta=\ln(r/r^{*})$, Eqs. (\ref{eq:1st})-(\ref{eq:4th}) now become
	\begin{equation}\begin{split}\label{eq:5th}
	\ddot{F}-\left(1+\dfrac{1}{2}(\dot{\nu}\right.&+\left.\dot{\lambda})\right)\dot{F}+\big(\dot{\nu}+\dot{\lambda}\big)F+\dfrac{3}{4}\kappa ~\text{e}^{-\nu}\rho=0~,
	\end{split}\end{equation}
	\vspace{-2em}
	\begin{equation}%\left.
	\begin{split}\label{eq:6th}
	\ddot{\nu}-\dot{\nu}+\dot{\nu}^2-\dfrac{1}{2}&\left(\dot{\nu}+2\right)(\dot{\nu}+\dot{\lambda})%\\&
	-2\left(1-e^\lambda\right)=2\dfrac{\ddot{F}}{F}-\left(\dot{\lambda}+4\right)\dfrac{\dot{F}}{F}~,
	\end{split}%\right\}
	\end{equation}
	\vspace{-1.0em}
	\begin{equation}\label{eq:7th}
	\begin{array}{rcl}
	f=F\mathnormal{e}^{-\lambda-2\eta}\left[\ddot{\nu}-\dot{\nu}-\dfrac{1}{2}\left(\dot{\nu}+\dot{\lambda}\right)\dot{\nu}\right.%\\
	-2\dot{\lambda}+\left.\left(\dot{\nu}+4\right)\dfrac{\dot{F}}{F}\right],
	\end{array}
	\end{equation}
	\vspace{-1.0em}
	\begin{equation}\begin{split}\label{eq:8th}
	\mathcal{R}=2\dfrac{f}{F}-\dfrac{\kappa}{F} ~ \text{e}^{-\nu}\rho&-3\mathnormal{e}^{-\lambda-2\eta}\left[ \dfrac{\ddot{F}}{F}\right.-\dfrac{\dot{F}}{F}
	+\left( \dfrac{1}{2}(\dot{\nu}-\dot{\lambda})+2 \right)\left.\dfrac{\dot{F}}{F}\right],
	\end{split}\end{equation}
	where dot ($.$) denotes derivative with respect to $\eta$. In the vacuum ($\rho=0$), with the introduction $1/F\left(\mathrm{d}F/\mathrm{d}\eta\right)=u$, Eqs. (\ref{eq:5th}) and (\ref{eq:6th}) simplify to
	\begin{equation}\label{eq:eq7}
	\dot{u}+u^2-\left[1+\dfrac{1}{2}(\dot{\nu}+\dot{\lambda})\right]u+\dot{\nu}+\dot{\lambda}=0~,
	\end{equation}
	\vspace{-2.0em}
	\begin{equation}%\left.
	\begin{split}\label{eq:eq8}
	\ddot{\nu}-\dot{\nu}+\dot{\nu}^2&-\dfrac{1}{2}\left(\dot{\nu}+2\right)(\dot{\nu}+\dot{\lambda})%\\&
	-2\left(1-\mathnormal{e}^{\lambda}\right)=2\left(\dot{u}+u^2\right)-(\dot{\lambda}+4)u~.
	\end{split}%\right\}
	\end{equation}
	%Even though above equations are second order differential equations in $\nu$ but since $\nu$ has already been given as a function of $r$ earlier%after Eq. \ref{eq:metric}
	%, the introduction of $u$ makes Eqs. (\ref{eq:5th}) and (\ref{eq:6th}) first order with the increase in number of equations. 
	%Using ${\mathrm{d}u}/{\mathrm{d}\eta}+u^2$ from
	Solving Eq. (\ref{eq:eq8}) and Eq. (\ref{eq:eq7}), we get
	\begin{equation}%\left.
	\begin{split}\label{eq:eq12->11}
	\ddot{\nu}-\dot{\nu}+\dot{\nu}^2-\left(1+\dfrac{\dot{\nu}}{2}\right)&(\dot{\nu}+\dot{\lambda})-2\left(1-\mathnormal{e}^{\lambda}\right)%\\&
	+2\left(1-\dfrac{\dot{\nu}}{2}\right)u+2(\dot{\nu}+\dot{\lambda})=0.
	\end{split}%\right\}
	\end{equation}
	The set of equations \{\ref{eq:eq7}~,~\ref{eq:eq12->11}\} or \{\ref{eq:eq8}~,~\ref{eq:eq12->11}\} provide the solution of $\lambda$ and $u$ ($\nu$ is already known).
	%%%%%%%%%%%%%%%%%%%%%%%%%%%%%%%%%%%%%%%%%%%%%%%%%%%%%%%%%%%%%%%%%%%%%%%%%%%%%%%%%%%%%%%%%
	\section{%Solution to the $f(\mathcal{R})$ gravity vacuum field equations to explain 
		Galactic rotational curves as a geometric effect in vacuum}\label{darkmatter}
	To obtain the geometric interpretation of the constant velocity regions, we solve Eqs. (\ref{eq:eq7}) and (\ref{eq:eq12->11}) for $\lambda$ and $u$ with $\nu=2m(\eta-\eta_0)$. Eqs. (\ref{eq:eq7}) and (\ref{eq:eq12->11}) can now be written as
	\begin{equation}\label{eq:DMeq1}
	\dot{u}+u^2-\left(1+m+\dfrac{\dot{\lambda}}{2}\right)u+2m+\dot{\lambda}=0~,
	\end{equation}
	\vspace{-2.0em}
	\begin{equation}\begin{split}\label{eq:DMeq2}
	-2m+4m^2-\left(1+m\right)\left(2m+\dot{\lambda}\right)&-2\left(1-\mathnormal{e}^{\lambda}\right)+2\left(1-m\right)u+2\left(2m+\dot{\lambda}\right)=0~.
	\end{split}\end{equation}
	Using $u$ and $\dot{u}$ from Eq. (\ref{eq:DMeq2}) into Eq. (\ref{eq:DMeq1}) to decouple the above equations, we obtain
	\begin{align}\begin{split}\label{eq:DMeq3}
	\dfrac{\ddot{\lambda}}{2}+\dfrac{1}{1-m}\left(\dfrac{3}{2}\right.\mathnormal{e}^\lambda&-m(1-m)\bigg)\dot{\lambda}-\dfrac{\dot{\lambda}^2}{2}-\dfrac{(1-\mathnormal{e}^\lambda)^2}{(1-m)^2}-\dfrac{\mathnormal{e}^\lambda(1+m^2)-1}{(1-m)^2}+2m=0.
	\end{split}\end{align}
	This is a second order differential equation in $\lambda$. After obtaining $\lambda$ from this equation one can use either Eq. (\ref{eq:DMeq1}) or Eq. (\ref{eq:DMeq2}) to obtain $u$. %The obtained $\lambda$ and $u$ must satisfy the third equation (which has not been used in getting $\lambda$ and $u$).
	As discussed in \citet{Bohmer2008386}, the tangential velocity of a test particle in a stable circular orbit around the galactic center is about $200-300\text{Km/s}$ \citet{Salucci11062007,Persic01071996,Borriello11052001} thus $m\approx\mathcal{O}(10^{-6})$ \footnote{The symbol $\mathcal{O}$ represents order of meaning between $1\times 10^{-6}$ to $9.9\times 10^{-6}$}. Hence, we should get an approximate function of Ricci scalar around $m\approx\mathcal{O}(10^{-6})$.
	
\begin{figure}%[h!]
		\centerline{
			\includegraphics[scale=0.5]{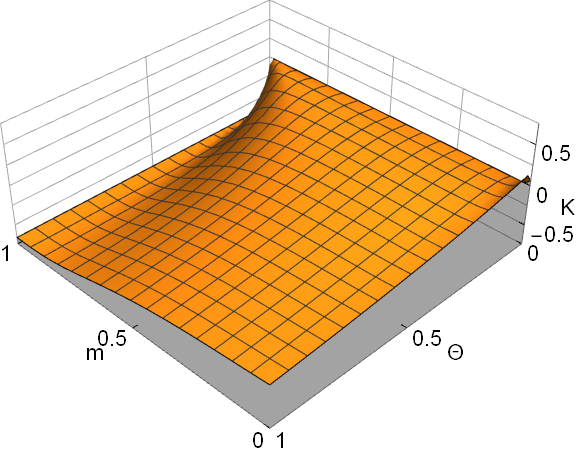}}
		\caption[K as a function of $\Theta$ and $m$.]{K as a function of $\Theta$ and $m$.}
		\label{fig:theta0}
	\end{figure}

	Introducing a new variable as $\mathrm{d}\lambda/\mathrm{d}\eta=\rho$ then we can write $\mathrm{d}/\mathrm{d}\eta=\rho~\mathrm{d}/\mathrm{d}\lambda$. This allows us to express Eq. (\ref{eq:DMeq3}) as
	\begin{align}\begin{split}\label{eq:DMnewvarible}
	&\hspace{-0.2cm}\dfrac{\rho}{2}~\dfrac{\mathrm{d}\rho}{\mathrm{d}\lambda}+\dfrac{1}{1-m}\left(\dfrac{3}{2}\right.\mathnormal{e}^\lambda-m(1-m){\bigg)}\rho-\dfrac{1}{2}\rho^2 -\dfrac{(1-\mathnormal{e}^\lambda)^2}{(1-m)^2}-\dfrac{\mathnormal{e}^\lambda(1+m^2)-1}{(1-m)^2}+2m=0.
	\end{split}\end{align}
	Now, using the transformations $\rho=1/\omega$ and $e^{\lambda}=\Theta$, Eq. (\ref{eq:DMnewvarible}) becomes
	\begin{align}\begin{split}\label{eq:DMtransformed}
	\dfrac{\mathrm{d}\omega}{\mathrm{d}\Theta}=\left[ \dfrac{m}{\Theta}\right.&-\dfrac{2(1-\Theta)^2}{\Theta(1-m)^2}-\left.\dfrac{2(\Theta(1+m^2)-1)}{\Theta(1-m)^2} \right]\omega^3%\\&
	+\left[ 3-\dfrac{2m(1-m)}{\Theta} \right]\omega^2-\dfrac{1}{\Theta}\omega~,
	\end{split}\end{align}
	which is a nonlinear first order Abel differential equation of the first kind of the form
	\begin{equation*}
	\dfrac{\mathrm{d}\omega}{\mathrm{d}\Theta}=p(\Theta)\omega^3+q(\Theta)\omega^2+r(\Theta)\omega+s(\Theta)~,
	\end{equation*}
	where
	\begin{equation*}
	p(\Theta)=\dfrac{m}{\Theta}-\dfrac{2(1-\Theta)^2}{\Theta(1-m)^2}-\dfrac{2(\Theta(1+m^2)-1)}{\Theta(1-m)^2}~,
	\end{equation*}
	\begin{equation*}
	q(\Theta)=3-\dfrac{2m(1-m)}{\Theta} \text{\space , \qquad\quad} r(\Theta)=-\dfrac{1}{\Theta} ~\text{\space , \qquad\quad} s(\Theta)=0.
	\end{equation*}
	If we choose that the particular solution $y_p$ of the above equation satisfies
	\begin{equation}\begin{split}\label{transformationcondition}
	3\left[ \dfrac{m}{\Theta}\right.&-\dfrac{2(1-\Theta)^2}{\Theta(1-m)^2}-\left.\dfrac{2(\Theta(1+m^2)-1)}{\Theta(1-m)^2} \right] y_p^2 %\\&
	+2\left[ 3-\dfrac{2m(1-m)}{\Theta} \right] y_p-\dfrac{1}{\Theta}=0~,
	\end{split}\end{equation}
	then Eq. (\ref{eq:DMtransformed}) becomes Abel differential equation of second kind \citet{Mak200291}, of the form
	\begin{align}\begin{split}\label{eq:DMtransformedAbel2nd}
	\dfrac{\mathrm{d}\omega}{\mathrm{d}\Theta}=&\left[ \dfrac{m}{\Theta}-\dfrac{2(1-\Theta)^2}{\Theta(1-m)^2}-\dfrac{2(\Theta(1+m^2)-1)}{\Theta(1-m)^2} \right]\omega^3 \\&
	\text{\qquad\qquad\qquad\qquad\qquad}+\left[3\left(\dfrac{m}{\Theta}-\dfrac{2(1-\Theta)^2}{\Theta(1-m)^2}-\dfrac{2(\Theta(1+m^2)-1)}{\Theta(1-m)^2}\right)y_p\right.%\\&~~~
	+\left.\left(3-\dfrac{2m(1-m)}{\Theta}\right) \right]\omega^2 ~.
	\end{split}\end{align}
	%In the case when,
	Also when we have,
	\begin{equation}\label{solutioncondition}
	\dfrac{\mathrm{d}}{\mathrm{d}\Theta}\left(\dfrac{p}{\sqrt{q^2-3pr}}\right)=K\sqrt{q^2-3pr}~,
	\end{equation}
	where $K$ is an arbitrary constant. The general solution of Eq. (\ref{eq:DMtransformed}) is given by \citet{Mak200291},
	\begin{equation}
	\begin{split}
	\omega(\Theta)=&\pm\dfrac{\sqrt{q^2-3pr}}{p}V%\\&
	+\left(\dfrac{-q\pm\sqrt{q^2-3pr}}{3p}\right),
	\end{split}
	\end{equation}
	where $V$ satisfies the separable differential equation
	\begin{equation}
	\dfrac{\mathrm{d}V}{\mathrm{d}\Theta}=\left(V^3+V^2+K~V\right)\left(\dfrac{q^2-3pr}{p}\right).
	\end{equation}
	Thus to get the solution, we must satisfy the condition given by Eq. (\ref{solutioncondition}). The plot of 
	\begin{equation*}
	K=\dfrac{1}{\sqrt{q^2-3pr}}~{\dfrac{\mathrm{d}}{\mathrm{d}\Theta}\left(\dfrac{p}{\sqrt{q^2-3pr}}\right)}
	\end{equation*} 
	is given in Fig. (\ref{fig:theta0}). Here, $p$, $q$ and $r$ are the functions of $\Theta$ as defined earlier.
	The Fig. (\ref{fig:theta0}) allows us to understand when (i.e. for what range of $m$ and $\Theta$) %%%When $\Theta=0$ we have Eq. (\ref{solutioncondition}) satisfied for approximately all values of $m={v^2}/{c^2}$. This demands an arbitrary negative large value of $\lambda(r)$. 
	when we have $m=1$ Eq. (\ref{solutioncondition}) is obeyed for all values of $\Theta$ except $-\infty$. For all other values of $m$ if $0\leq\Theta\leq1$ (i.e. all possible values, ($0,\infty$) is being mapped onto ($0,1$) with the mapping function $e^{-1/x}$) then we observe the violation of Eq. (\ref{solutioncondition}) (i.e. $K$ is not a constant).
	%It has also be proven in \citet{Usman2016} that $\lambda=\text{constant}$ is not the solution of Eq. (\ref{eq:DMeq1}) and (\ref{eq:DMeq2}) in the limit that $m^2$ and higher order terms in $m$ can be neglected.
	%%%%%%%%%%%%%%%%%%%%%%%%%%%%%%%%%%%%%%%%%%%%%%%%%%%%%%%%%%%%%%%%%%%%%%%%%%%%%%%%%%%%%%%%%
	\section{Numerical solution to the field equations}\label{numericalresults}
	
\begin{figure}%[h!]
		\begin{minipage}{0.47\linewidth}
			\centering
			\includegraphics[width=0.9\textwidth]{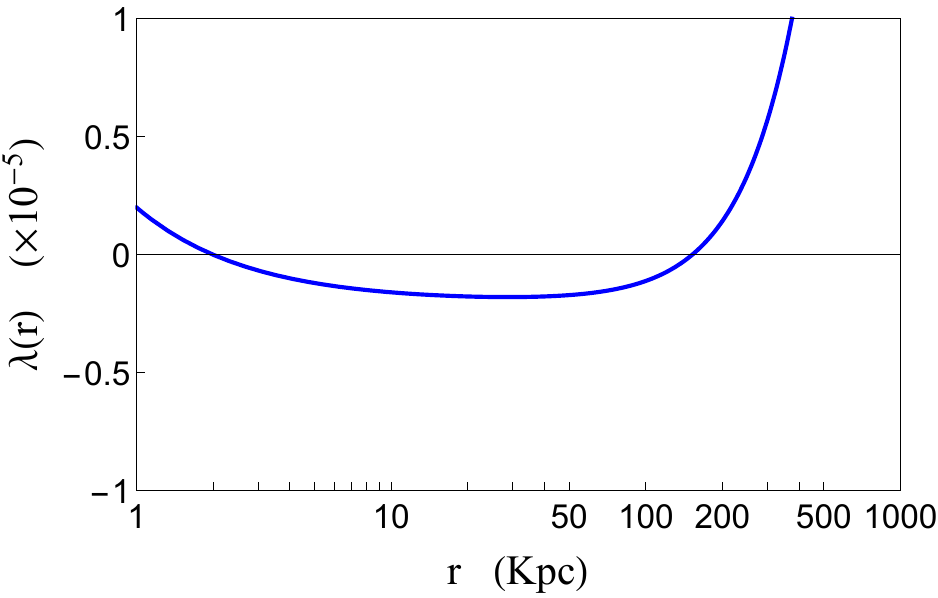}
			\subcaption{For vacuum, $\rho=0$}
			\label{fig:Lambda-sub1}
		\end{minipage}\hspace{0.5cm}
		\begin{minipage}{0.47\linewidth}
			\centering
			\includegraphics[width=0.9\textwidth]{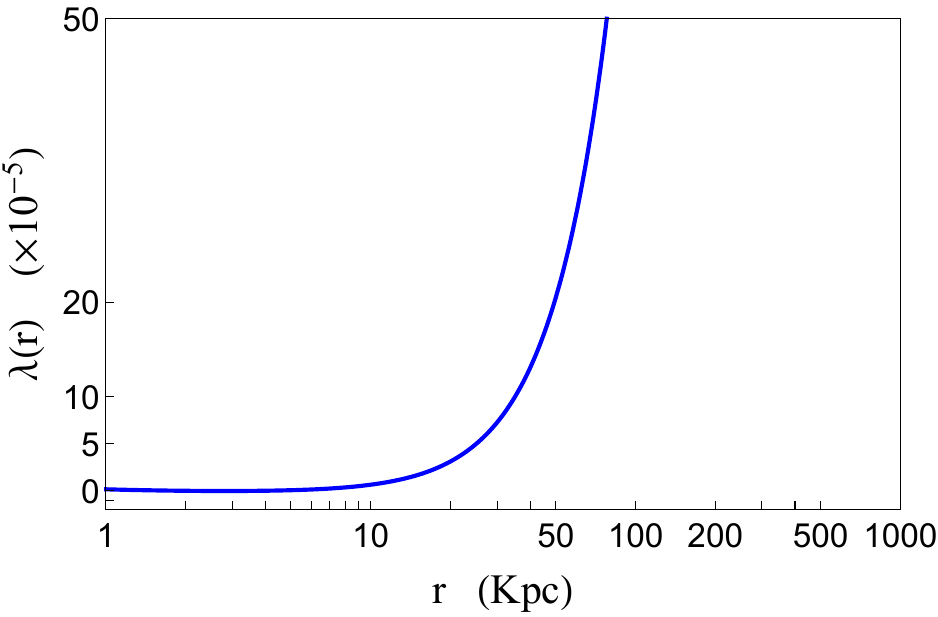}
			\subcaption{For matter density profile given by Eq. (\ref{eq:Power-law-density-profile})}
			\label{fig:Lambda-sub2}
		\end{minipage}\\[1ex]
		\begin{minipage}{0.47\linewidth}
			\centering
			\includegraphics[width=0.9\textwidth]{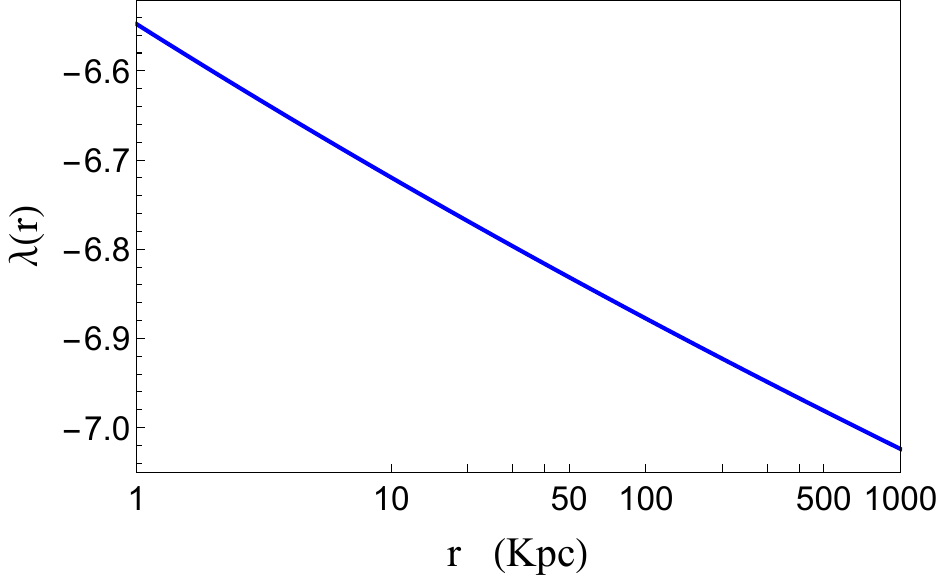}
			\subcaption{For matter density profile given by Eq. (\ref{eq:Simple-model-for-a-galaxy-with-a-core})}
			\label{fig:Lambda-sub3}
		\end{minipage}\hspace{0.5cm}
		\begin{minipage}{0.47\linewidth}
			\centering
			\includegraphics[width=0.9\textwidth]{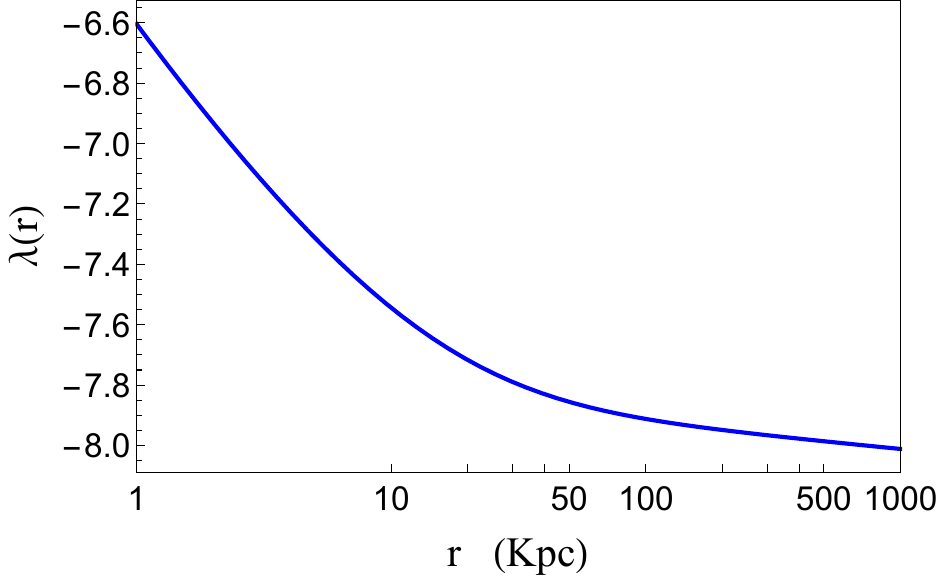}
			\subcaption{M31 galaxy, For matter density profile given by Eq. (\ref{eq:NFW-profile})}
			\label{fig:Lambda-sub4}
		\end{minipage}\\[1ex]
		\begin{minipage}{0.47\linewidth}
			\centering
			\includegraphics[width=0.9\textwidth]{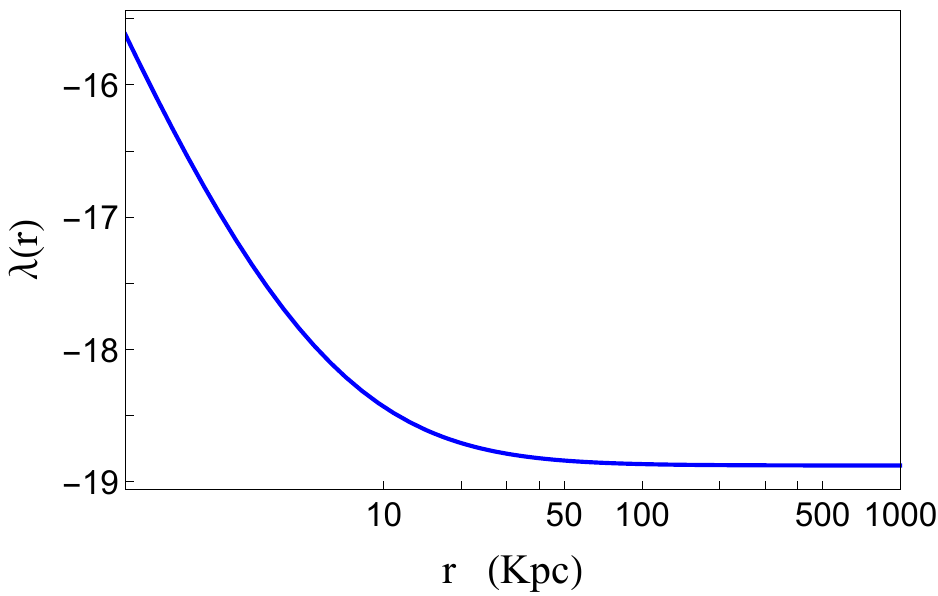}
			\subcaption{Millky Way galaxy, For matter density profile given by Eq. (\ref{eq:NFW-profile})}
			\label{fig:Lambda-sub5}
		\end{minipage}\hspace{0.5cm}
		\begin{minipage}{0.47\linewidth}
			\centering
			\includegraphics[width=0.9\textwidth]{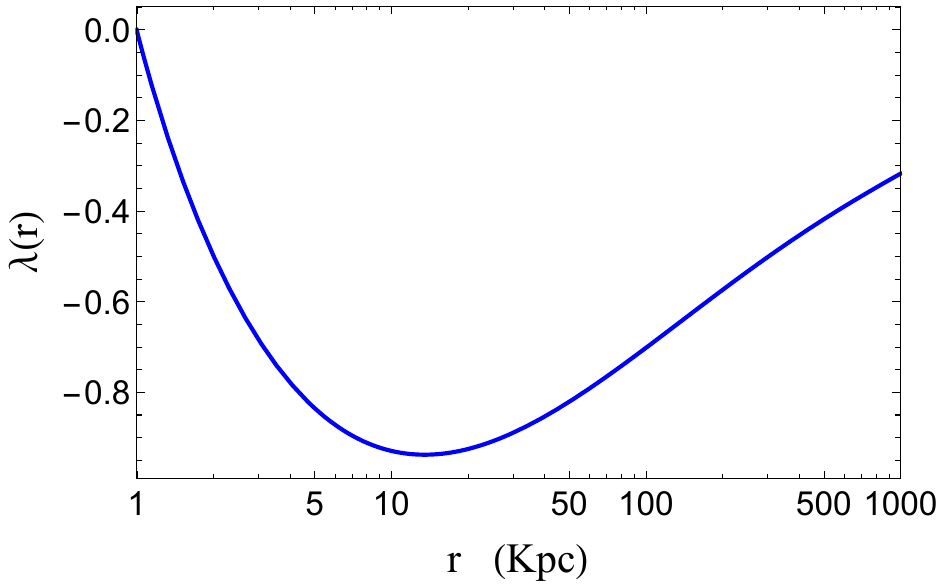}
			\subcaption{For matter density profile given by Eq. (\ref{eq:Power-law-density-profile}) with $\rho_0r_0^2=1.05792\times10^{45}\text{Kg/m}$}
			\label{fig:Lambda-sub6}
		\end{minipage}
		\caption{The radial component $\lambda(r)$ of the metric for different galactic matter density profiles.}
		\label{fig:Lambda}
	\end{figure}

\begin{figure}%[h!]
		\begin{minipage}{0.47\linewidth}
			\centering
			\includegraphics[width=0.9\textwidth]{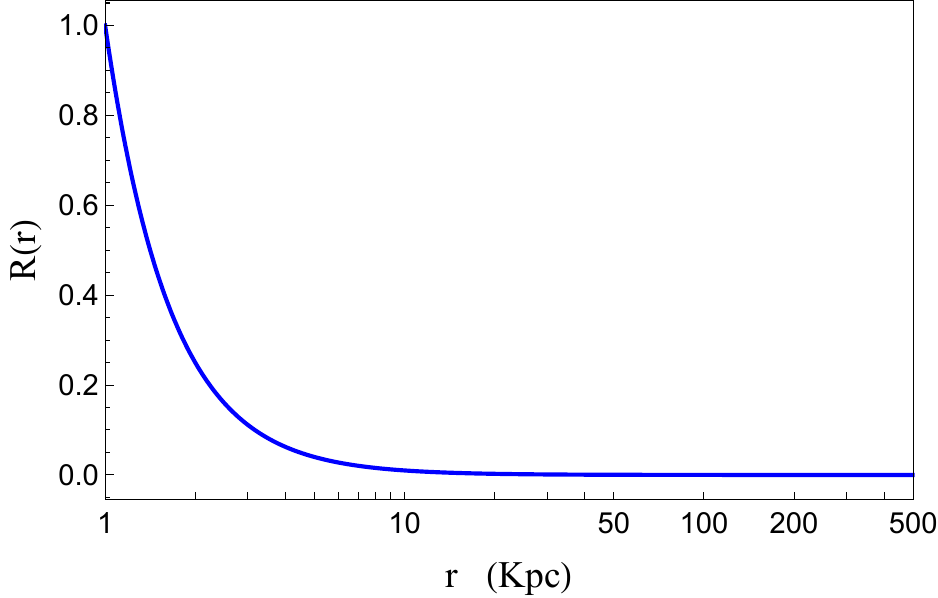}
			\subcaption{For vacuum, $\rho=0$}
			\label{fig:Ricci(r)-sub1}
		\end{minipage}\hspace{0.5cm}
		\begin{minipage}{0.47\linewidth}
			\centering
			\includegraphics[width=0.9\textwidth]{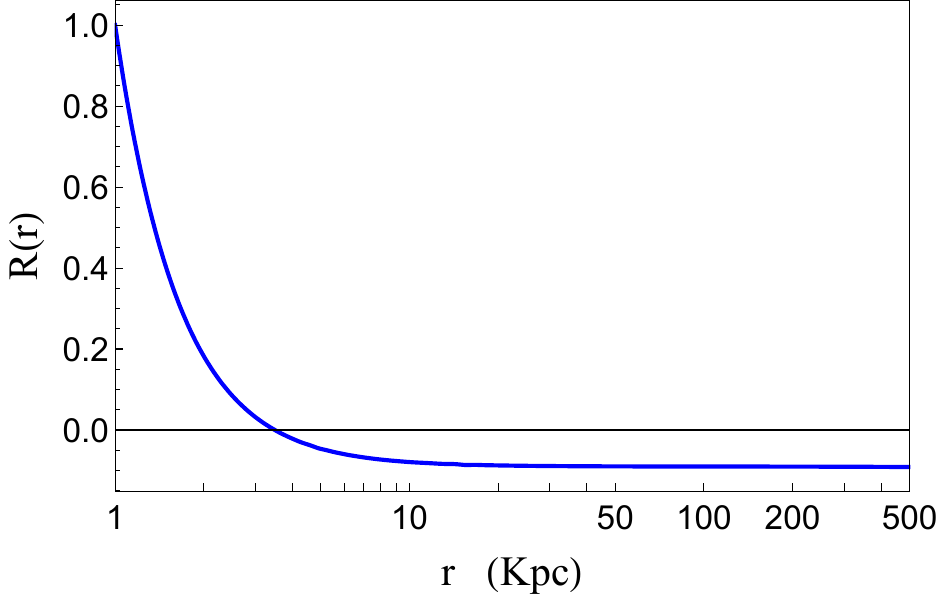}
			\subcaption{For matter density profile given by Eq. (\ref{eq:Power-law-density-profile})}
			\label{fig:Ricci(r)-sub2}
		\end{minipage}\\[1ex]
		\begin{minipage}{0.47\linewidth}
			\centering
			\includegraphics[width=0.9\textwidth]{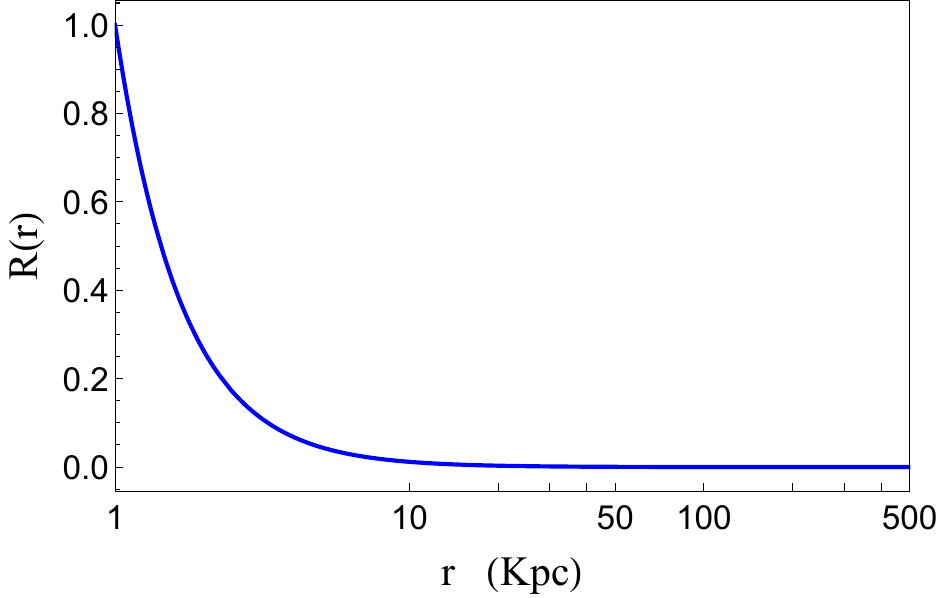}
			\subcaption{For matter density profile given by Eq. (\ref{eq:Simple-model-for-a-galaxy-with-a-core})}
			\label{fig:Ricci(r)-sub3}
		\end{minipage}\hspace{0.5cm}
		\begin{minipage}{0.47\linewidth}
			\centering
			\includegraphics[width=0.9\textwidth]{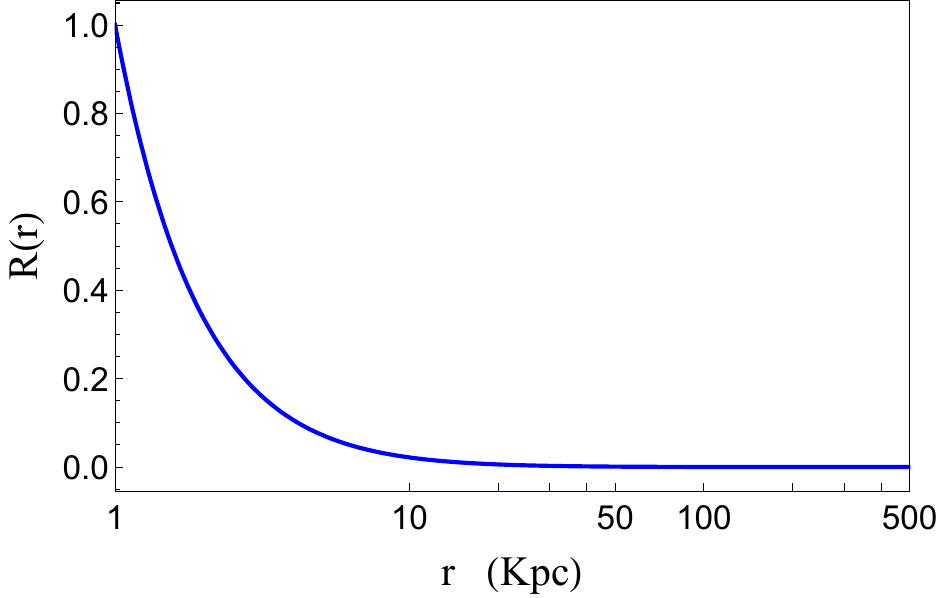}
			\subcaption{M31 galaxy, For matter density profile given by Eq. (\ref{eq:NFW-profile})}
			\label{fig:Ricci(r)-sub4}
		\end{minipage}\\[1ex]
		\begin{minipage}{0.47\linewidth}
			\centering
			\includegraphics[width=0.9\textwidth]{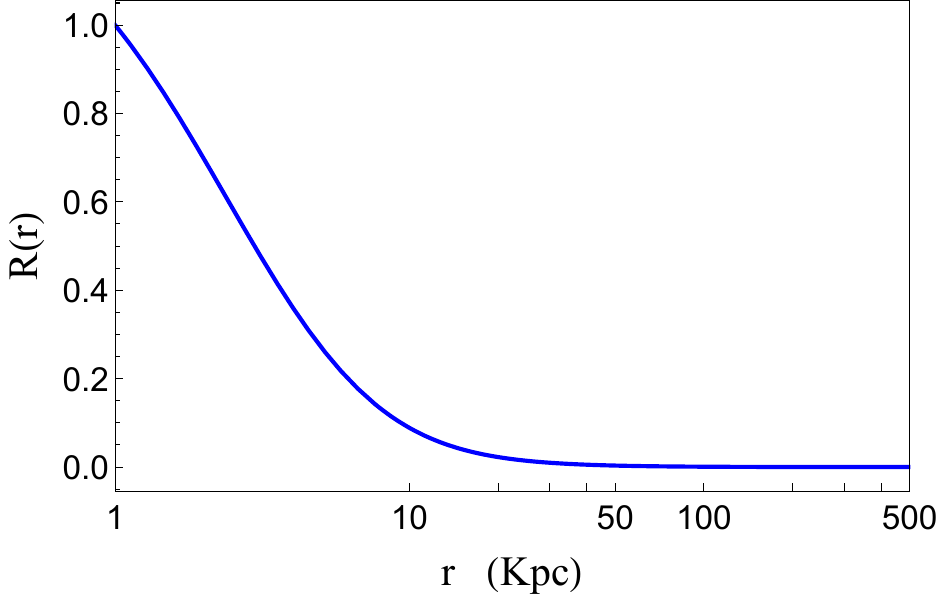}
			\subcaption{Millky Way galaxy, For matter density profile given by Eq. (\ref{eq:NFW-profile})}
			\label{fig:Ricci(r)-sub5}
		\end{minipage}\hspace{0.5cm}
		\begin{minipage}{0.47\linewidth}
			\centering
			\includegraphics[width=0.9\textwidth]{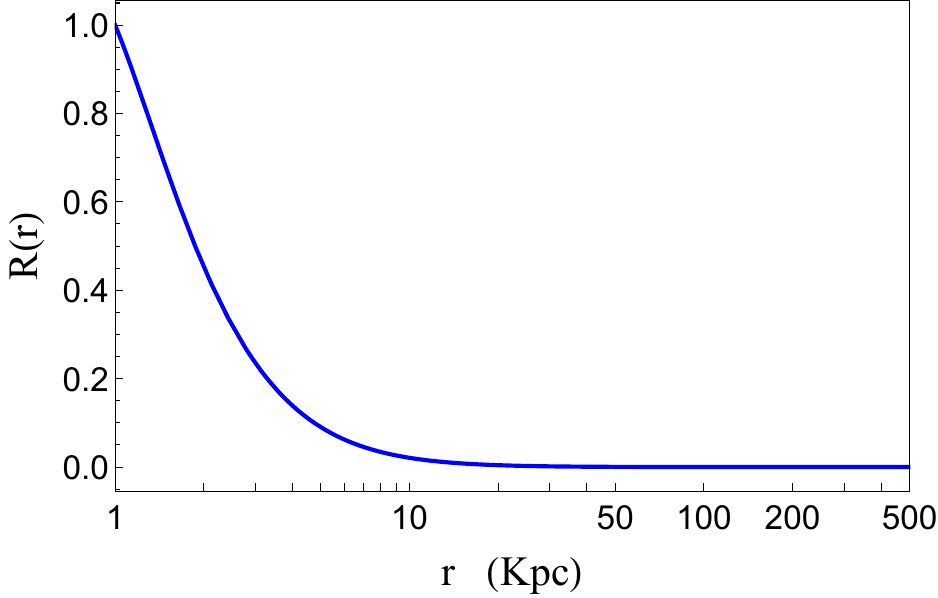}
			\subcaption{For matter density profile given by Eq. (\ref{eq:Power-law-density-profile}) with $\rho_0r_0^2=1.05792\times10^{45}\text{Kg/m}$}
			\label{fig:Ricci(r)-sub6}
		\end{minipage}
		\caption[Normalized $\mathcal{R}$ as a function of $r$]{The Normalized Ricci scalar $\mathcal{R}_N$ (defined as  $\mathcal{R}_N\equiv\mathnormal{\mathcal{R}}/\mathcal{R}_{ini}$) as a function of the radial parameter $r$, the distance from the galactic center.}
		\label{fig:Ricci(r)}
	\end{figure}

\begin{figure}%[h!]
		\begin{minipage}{0.47\linewidth}
			\centering
			\includegraphics[width=0.9\textwidth]{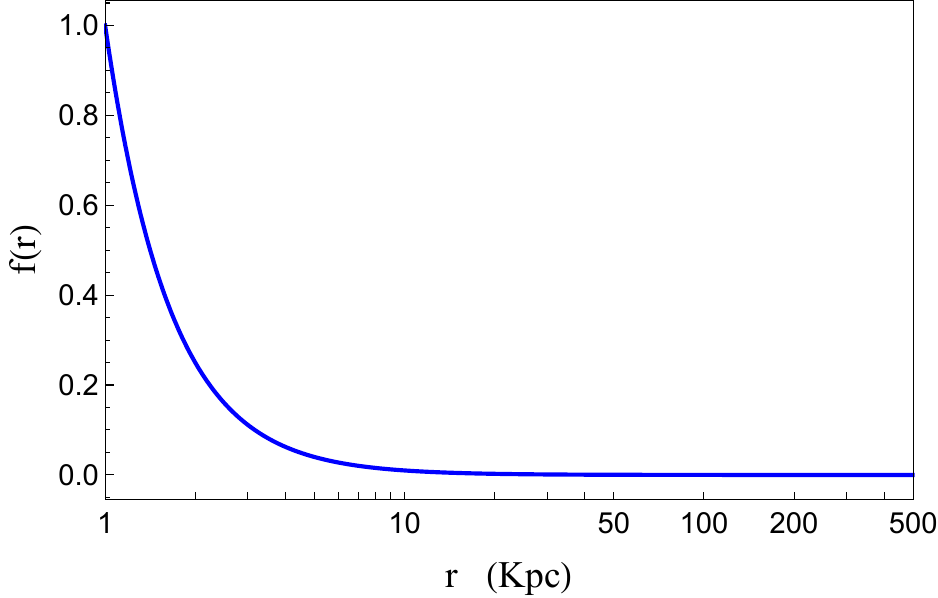}
			\subcaption{For vacuum, $\rho=0$}
			\label{fig:fr-sub1}
		\end{minipage}\hspace{0.5cm}
		\begin{minipage}{0.47\linewidth}
			\centering
			\includegraphics[width=0.9\textwidth]{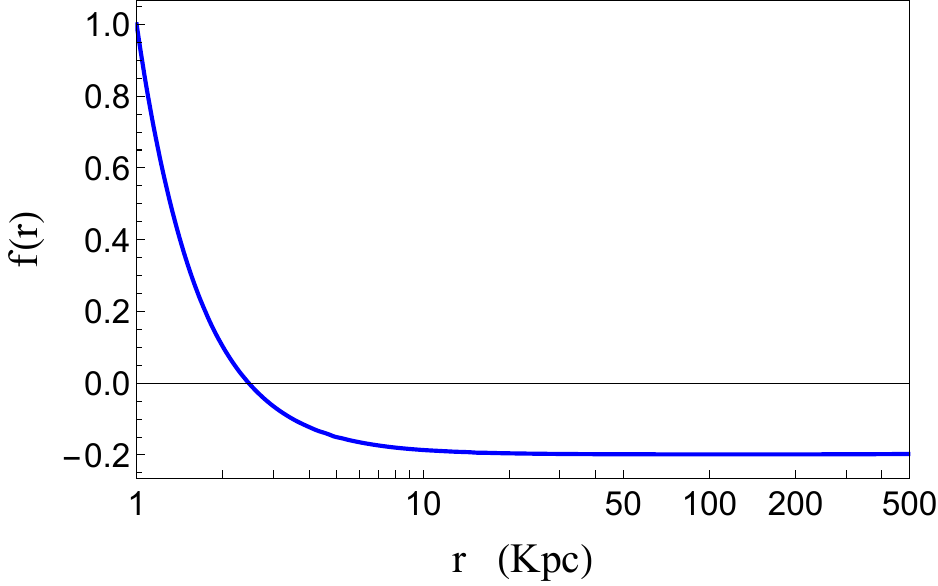}
			\subcaption{For matter density profile given by Eq. (\ref{eq:Power-law-density-profile})}
			\label{fig:fr-sub2}
		\end{minipage}\\[1ex]
		\begin{minipage}{0.47\linewidth}
			\centering
			\includegraphics[width=0.9\textwidth]{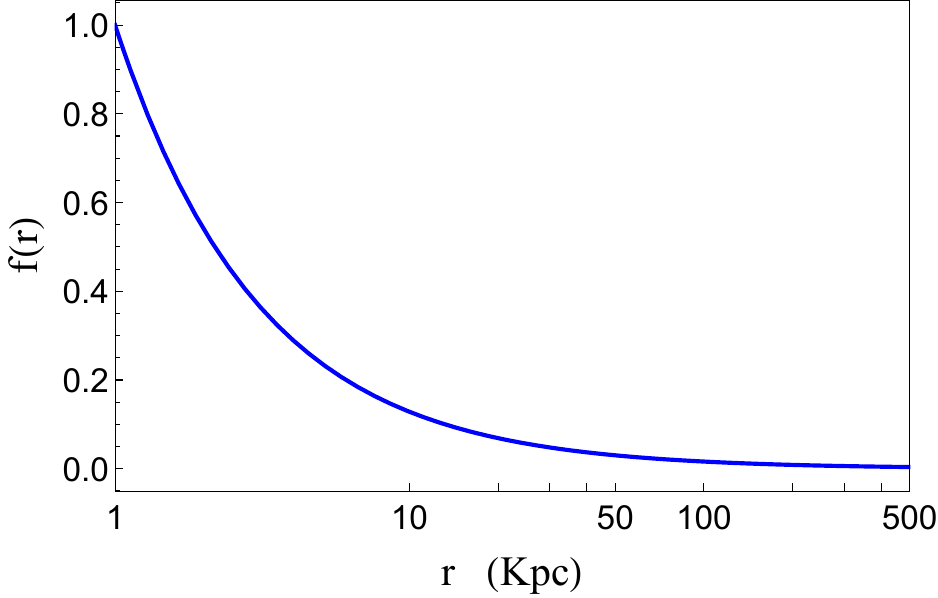}
			\subcaption{For matter density profile given by Eq. (\ref{eq:Simple-model-for-a-galaxy-with-a-core})}
			\label{fig:fr-sub3}
		\end{minipage}\hspace{0.5cm}
		\begin{minipage}{0.47\linewidth}
			\centering
			\includegraphics[width=0.9\textwidth]{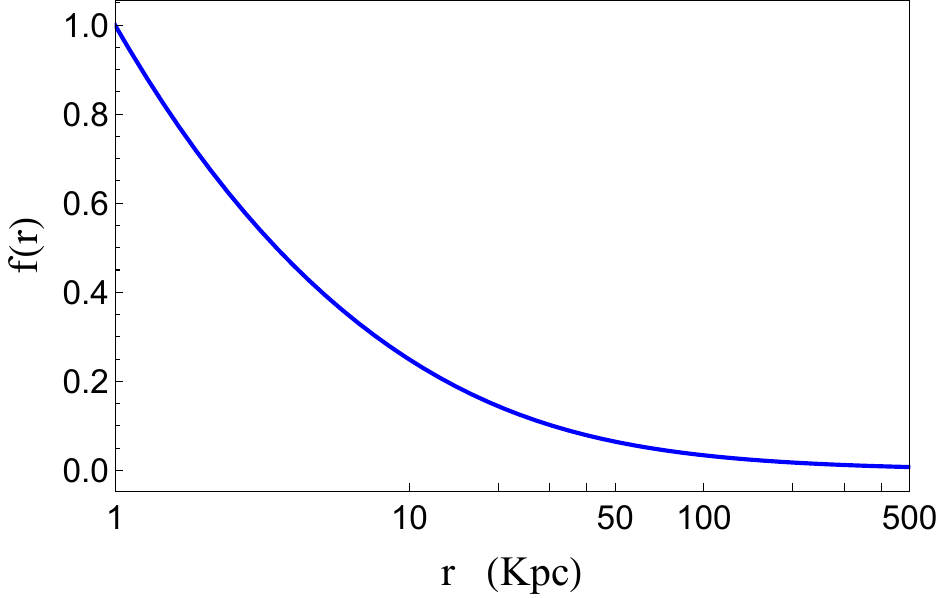}
			\subcaption{M31 galaxy, For matter density profile given by Eq. (\ref{eq:NFW-profile})}
			\label{fig:fr-sub4}
		\end{minipage}\\[1ex]
		\begin{minipage}{0.47\linewidth}
			\centering
			\includegraphics[width=0.9\textwidth]{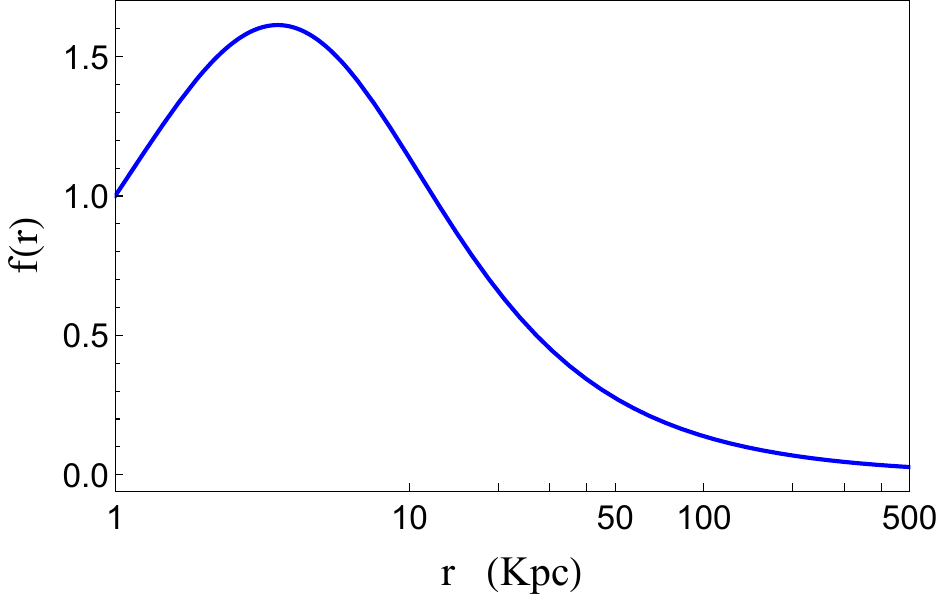}
			\subcaption{Millky Way galaxy, For matter density profile given by Eq. (\ref{eq:NFW-profile})}
			\label{fig:fr-sub5}
		\end{minipage}\hspace{0.5cm}
		\begin{minipage}{0.47\linewidth}
			\centering
			\includegraphics[width=0.9\textwidth]{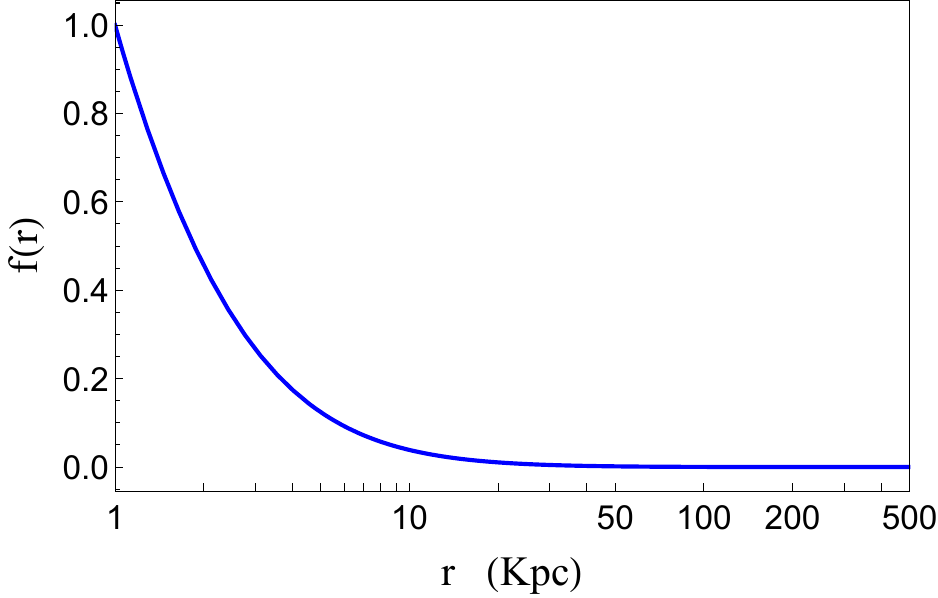}
			\subcaption{For matter density profile given by Eq. (\ref{eq:Power-law-density-profile}) with $\rho_0r_0^2=1.05792\times10^{45}\text{Kg/m}$}
			\label{fig:fr-sub6}
		\end{minipage}
		\caption[Normalized $f$ as a function of $r$]{The Normalized modified gravity function $f_N$ (defined as  $f_N\equiv\mathnormal{f}/\mathnormal{f}_{ini}$) as a function of the radial parameter $r$, the distance from the galactic center.}
		\label{fig:fr}
	\end{figure}

\begin{figure}%[h!]
		\begin{minipage}{0.47\linewidth}
			\centering
			\includegraphics[width=0.9\textwidth]{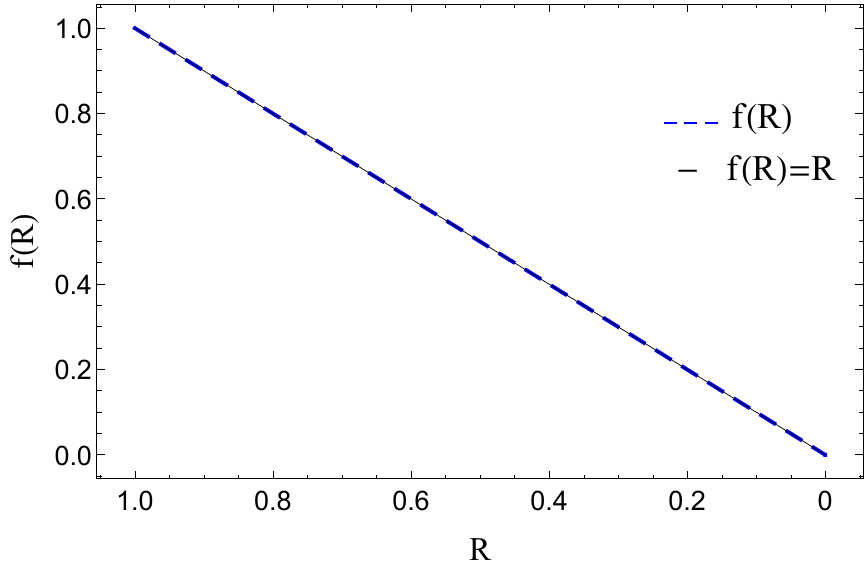}
			\subcaption{For vacuum, $\rho=0$}
			\label{fig:f(R)-sub1}
		\end{minipage}\hspace{0.5cm}
		\begin{minipage}{0.47\linewidth}
			\centering
			\includegraphics[width=0.9\textwidth]{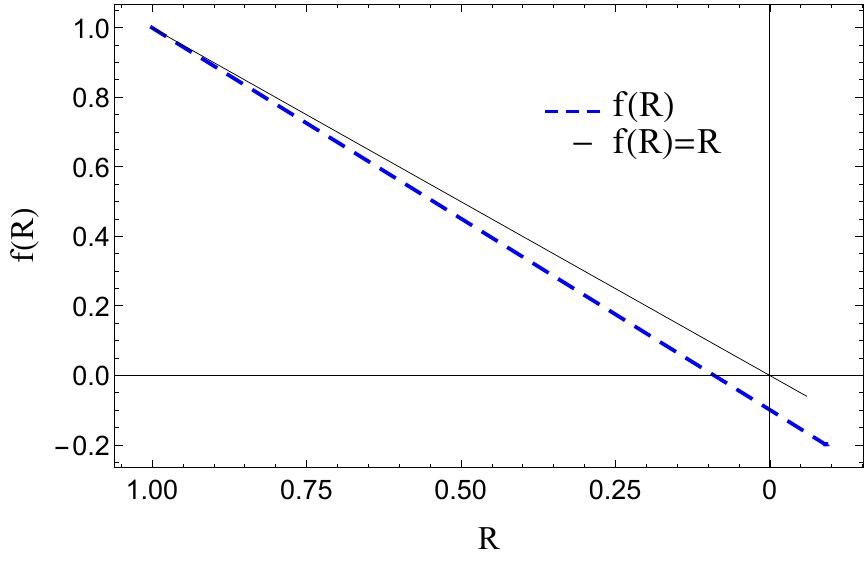}
			\subcaption{For matter density profile given by Eq. (\ref{eq:Power-law-density-profile})}
			\label{fig:f(R)-sub2}
		\end{minipage}\\[1ex]
		\begin{minipage}{0.47\linewidth}
			\centering
			\includegraphics[width=0.9\textwidth]{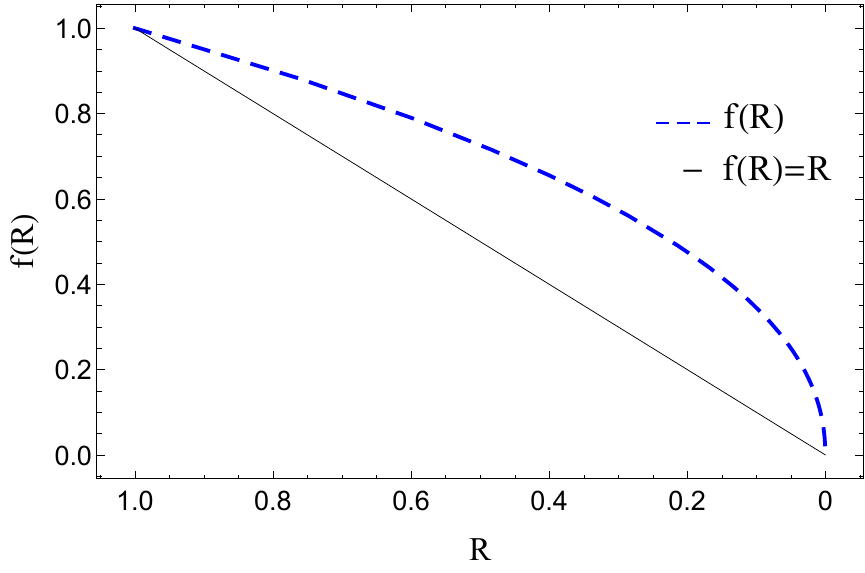}
			\subcaption{For matter density profile given by Eq. (\ref{eq:Simple-model-for-a-galaxy-with-a-core})}
			\label{fig:f(R)-sub3}
		\end{minipage}\hspace{0.5cm}
		\begin{minipage}{0.47\linewidth}
			\centering
			\includegraphics[width=0.9\textwidth]{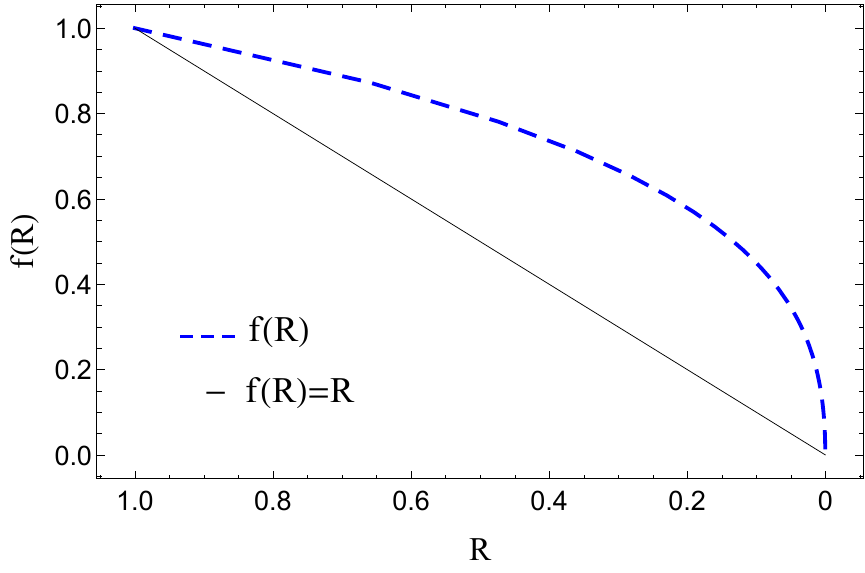}
			\subcaption{M31 galaxy, For matter density profile given by Eq. (\ref{eq:NFW-profile})}
			\label{fig:f(R)-sub4}
		\end{minipage}\\[1ex]
		\begin{minipage}{0.47\linewidth}
			\centering
			\includegraphics[width=0.9\textwidth]{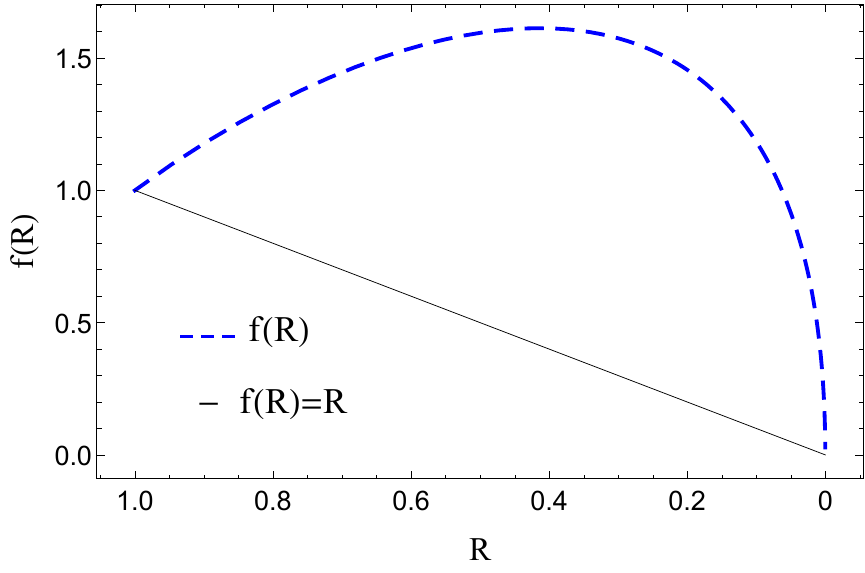}
			\subcaption{Millky Way galaxy, For matter density profile given by Eq. (\ref{eq:NFW-profile})}
			\label{fig:f(R)-sub5}
		\end{minipage}\hspace{0.5cm}
		\begin{minipage}{0.47\linewidth}
			\centering
			\includegraphics[width=0.9\textwidth]{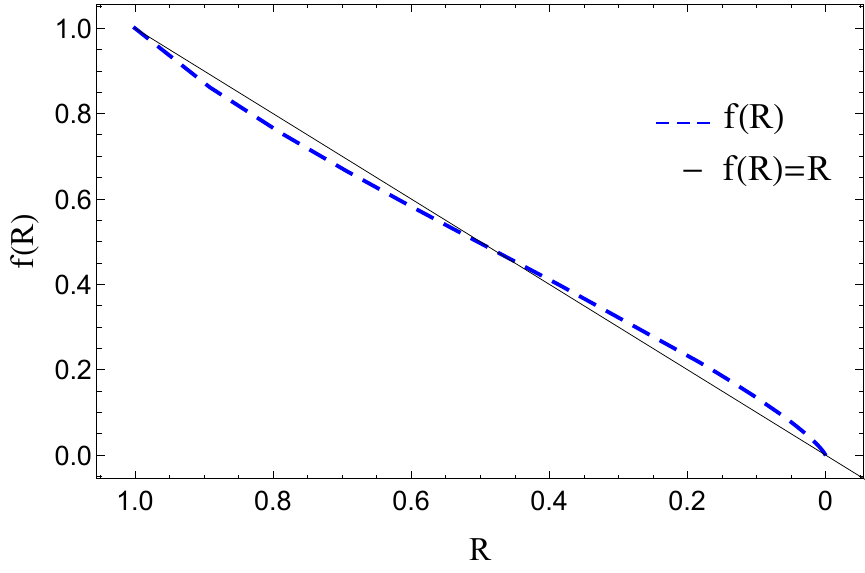}
			\subcaption{For matter density profile given by Eq. (\ref{eq:Power-law-density-profile}) with $\rho_0r_0^2=1.05792\times10^{45}\text{Kg/m}$}
			\label{fig:f(R)-sub6}
		\end{minipage}
		\caption{The modified gravity function as a function of Ricci scalar.}
		\label{fig:f(R)}
	\end{figure}

	From the Fig. (\ref{fig:theta0}), analytic solution of Eq. (\ref{eq:DMtransformed}) in the presented scheme around $m=\mathcal{O}(10^{-6})$ can not be obtained for all possible values of $\theta$ in vacuum ($\rho=0$). Since, with any form of the matter the $f(\mathcal{R})$ gravity equation have even complex form, the analytic solution for any matter distribution is out of question. We then find the numerical solution to the field Eqs. (\ref{eq:1st}) and (\ref{eq:2nd}) in vacuum as well as with the following density profiles corresponding to the galactic scenario. 
	\begin{enumerate}
		
		\item \textbf{Power law density profile:} If the galactic matter density as a function of $r$ is 
		\begin{equation}\label{eq:Power-law-density-profile}
		\rho(r)=\rho_0 \bigg(\dfrac{r_0}{r}\bigg)^2~,
		\end{equation}
		the corresponding tangential velocity becomes constant which is
		\begin{equation}\label{eq:velocity-for-Power-law-density-profile}
		V_t=\sqrt{\dfrac{4\pi G \rho_0 r_0^2}{3}}~.
		\end{equation}
		Since the observed tangential velocity around the galaxy at large distances is constant and is about $200-300\text{Km/s}$ \citet{Salucci11062007,Persic01071996,Borriello11052001}, this corresponds to, (with $V_t=300\text{Km/s}$), $\rho_0 r_0^2=1.073\times 10^{20}\text{Kg/m}$. We also obtain the numerical solution corresponding to $\rho_0r_0^2=1.05792\times10^{45}\text{Kg/m}$ to see the variational behavior of $f(\mathcal{R})$ for this type of matter distribution. The numerical solutions are presented in Figures (\ref{fig:Lambda-sub2}), (\ref{fig:Lambda-sub6}), (\ref{fig:Ricci(r)-sub2}), (\ref{fig:Ricci(r)-sub6}), (\ref{fig:fr-sub2}), (\ref{fig:fr-sub6}), (\ref{fig:f(R)-sub2}) and (\ref{fig:f(R)-sub6}).
		
		\item \textbf{Simple model for a galaxy with a core:} If we choose that the matter is distributed in the galaxy according to
		\begin{equation}\label{eq:Simple-model-for-a-galaxy-with-a-core}
		\rho(r)=\dfrac{1}{4\pi G} ~ \dfrac{V_t^2}{r^2+a_H^2},
		\end{equation}
		then the density at small distances, $r$, remains constant whereas it varies as $r^{-2}$ at large distances, $r$. This corresponds to the circular velocity curve
		\begin{equation}\label{eq:velocity-Simple-model-for-a-galaxy-with-a-core}
		V(r)=V_t ~ \sqrt{1-\dfrac{a_H}{r}\arctan{\bigg(\dfrac{r}{a_H}\bigg)}}~.
		\end{equation}
		The tangential velocity given by Eq. (\ref{eq:Simple-model-for-a-galaxy-with-a-core}) is plotted in Fig. (\ref{fig:vt}), which is very close to the observed behavior.
		\begin{figure}%[h!]
			\centering
			\includegraphics[width=10cm]{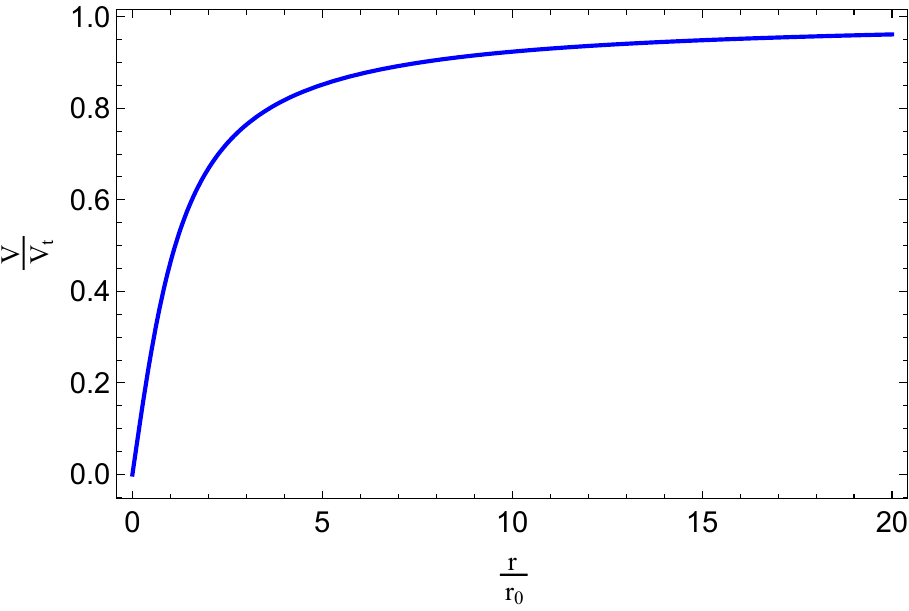}
			\caption[Velocity profile for simple model for a galaxy with a core.]{Velocity profile for simple model for a galaxy with a core.}
			\label{fig:vt}
		\end{figure}
		The numerical solution is given in Figures (\ref{fig:Lambda-sub3}), (\ref{fig:Ricci(r)-sub3}), (\ref{fig:fr-sub3}) and (\ref{fig:f(R)-sub3}) for $a_H=5\text{Kpc}$ and $V_t=300\text{Km/s}$.
		
		\item \textbf{Navarro, Frenk and White (NFW) profile:} In 1996, Julio F. Navarro, Carlos S. Frenk and Simon D. M. White gave a density profile which was fitted to dark matter haloes identified in N-body simulations \citet{1996ApJ...462..563N}. The NFW profile is
		\begin{equation}\label{eq:NFW-profile}
		\rho(r)=\dfrac{\rho_0}{\dfrac{r}{r_0}\left(1+\dfrac{r}{r_0}\right)^2}~.
		\end{equation}
		The free parameters $\rho_0$ and $r_0$ for Milky Way and M31 galaxy are given in Table. \ref{table:NFW-parameters}. %\citet{doi:10.1093/pasj/psv042},
		\begin{table}%[H]
			\begin{center}
				\begin{tabular}{|c|c|c|}
					\hline
					\textbf{Parameter} & \textbf{M31 galaxy} & \textbf{Milky Way galaxy} \\
					\hline\hline
					$r_0$(Kpc) & $34.6 \pm 2.1$ & $10.7 \pm 2.9$ \\ 
					\hline
					$\rho_0$ ($10^{-3}M_\odot \text{pc}^{-3}$) & $2.23\pm0.24$ & $18.2\pm7.4$ \\
					\hline
				\end{tabular}
			\end{center}
			\caption{The free parameters of NFW profile for M31 and Milky Way galaxy, \citet{doi:10.1093/pasj/psv042} .}
			\label{table:NFW-parameters}
		\end{table}
		
	\end{enumerate}
	\begin{itemize}
		\item The initial conditions for vacuum solution are chosen to be $r_{ini}=1~\text{Kpc}~, \text{\space}F(r_{ini})=1~,\text{\space} F'(r_{ini})=0~,\lambda(r_{ini})=2\times10^{-6}$ such that we have $f(\mathcal{R})=\mathcal{R}$ at $r=r_{ini}$.
		\item The initial conditions for matter density profile \ref{eq:Power-law-density-profile} are chosen to be $r_{ini}=1~\text{Kpc}~, \text{\space}F(r_{ini})=1~,\text{\space} F'(r_{ini})=0~,\lambda(r_{ini})=0.00000175018$ for plots given by Figs. (\ref{fig:Lambda-sub2}), (\ref{fig:Ricci(r)-sub2}), (\ref{fig:fr-sub2}) and (\ref{fig:f(R)-sub2}) %all (b) figures% 
		such that we have $f(\mathcal{R})=\mathcal{R}$ at $r=r_{ini}$, and $\lambda(r_{ini})=0.000001$ for Figs. (\ref{fig:Lambda-sub6}), (\ref{fig:Ricci(r)-sub6}), (\ref{fig:fr-sub6}) and (\ref{fig:f(R)-sub6}). %all (f) figures.
		\item For matter distribution given by Eq. (\ref{eq:Simple-model-for-a-galaxy-with-a-core}), initial conditions were chosen to be $r_{ini}=1~\text{m}~, \text{\space}F(r_{ini})=1~,\text{\space} F'(r_{ini})=0~,\lambda(r_{ini})=0$ for Figs. (\ref{fig:Lambda-sub3}), (\ref{fig:Ricci(r)-sub3}), (\ref{fig:fr-sub3}) and (\ref{fig:f(R)-sub3}).%all plots given by (c) figures.
		\item For NFW profiles, initial conditions were chosen to be $r_{ini}=1~\text{pc}~, \text{\space}F(r_{ini})=1~,\text{\space} F'(r_{ini})=0~,\lambda(r_{ini})=10$ for Figs. (\ref{fig:Lambda-sub4}), (\ref{fig:Ricci(r)-sub4}), (\ref{fig:fr-sub4}), (\ref{fig:f(R)-sub4}), (\ref{fig:Lambda-sub5}), (\ref{fig:Ricci(r)-sub5}), (\ref{fig:fr-sub5}) and (\ref{fig:f(R)-sub5}).%all plots given by (d) and (e) figures.
	\end{itemize}
	The initial conditions were chosen such that initially our model of $f(\mathcal{R})$ gravity matches onto GR (the value of $\lambda$ was chosen by brute force attack) for vacuum, power law density profile figures (b) and simple galactic core model. For NFW profile, initial conditions does not imply $\mathcal{R}_{ini}=f_{ini}$ but rather $\mathcal{R}_{ini}=\alpha f_{ini}$ where $\alpha$ is a arbitrary number which depends upon $\lambda_{ini}$.
	
	%\begin{figure}%[h!]
	%	\centerline{
	%		\includegraphics[width=12.5cm]{Lambda.pdf}}
	%	\caption[Metric coefficient $\lambda(r)$]{The radial coefficient of the metric. As we move away from the center of the galaxy $\lambda(r)$ decreases till $r\approx 300\text{ Kpc}$ then it starts to increase.}
	%	\label{fig:Lambda}
	%\end{figure}
	
	%\begin{figure*}[t!]
	%	$\begin{array}{rl}
	%	\includegraphics[width=0.5\textwidth]{Plots/Lambda.pdf} &
	%	\includegraphics[width=0.5\textwidth]{Plots/Lambda-power-law-density-profile.pdf}\\
	%	\includegraphics[width=0.5\textwidth]{Plots/Lambda-Simple-model-for-a-galaxy-with-a-core.pdf} &
	%	\includegraphics[width=0.5\textwidth]{Plots/Lambda-NFW-M31.pdf} \\ 
	%	\includegraphics[width=0.5\textwidth]{Plots/Lambda-NFW-Milky-Way.pdf} &
	%	\includegraphics[width=0.5\textwidth]{Plots/Lambda-rho_0=4,10^42.pdf}
	%	\end{array}$
	%	\caption[My beautiful figure.]{\label{fig:label}My beautiful figure}
	%\end{figure*}
	
	For the vaccum and power law density distribution ($\rho_0r_0^2=1.05792\times10^{45}\text{Kg/m}$, figures (f)), the metric coefficient $\lambda(r)$ decreases initially till approximately $r= 40\text{Kpc}$ and $r=12.6\text{Kpc}$ respectively, after that $\lambda$ is monotonically increasing function of distance.
	From the same Fig. (\ref{fig:Lambda}) we see that the metric coefficient $\lambda(r)$, for simple galactic model and NFW profiles, decreases and settles down to their respective minimum values.
	
	The Ricci scalar, $\mathcal{R}$, decreases as we move away from galactic center and becomes negligibly small between $100-500\text{Kpc}$ for all models except for power law density distribution ($\rho_0 r_0^2=1.073\times 10^{20}\text{Kg/m}$) where it permanently becomes negative. The smallness of the Ricci scalar shows that we have reached the galactic edge. As away from the galaxy, the volume of a geodesic sphere should match on to that of a ball in Euclidean space which is proved here as the Ricci scalar eventually becomes negligibly small. The negativity of the Ricci scalar for Fig. (\ref{fig:Ricci(r)-sub2}) is because of the continuous $r^{-2}$ dependence of the density from $r=1\text{Kpc}$ till the galactic edge. For this case, the Ricci scalar negative shows that the volume of a geodesic sphere is larger than the volume of a ball in Euclidean space because the continuous distribution of matter which distorts the space towards the point, such as being inside a spherical shell and hence we have the negative Ricci scalar.
	
	%\begin{figure}%[h!]
	%	\centerline{
	%		\includegraphics[width=12.5cm]{fr.pdf}}
	%	\caption[Normalized $f$ as a function of $r$]{Normalized $f$ (defined as  $f_N\equiv\mathnormal{f}/\mathnormal{f}_{ini}$) as a function of the radial parameter $r$, moving away from the center of the galaxy we have $f\rightarrow 0$.}
	%	\label{fig:f}
	%\end{figure}
	
	From Fig. (\ref{fig:fr}) we see that the normalized modified gravity function, $f(r)$, decreases monotonically as we go beyond the center of the galaxy thus matches with $\mathcal{R}=0$ case except for Fig. (\ref{fig:fr-sub2}) where it also gets negative just like $\mathcal{R}$ and we see a significant deviation from GR which becomes permanent even when we reach the galactic edge. 
	
	%\begin{figure}[h!p]
	%	\centerline{
	%		\includegraphics[width=12.5cm]{fRicci}}
	%	\caption[The $f\mathcal{R}$.]{The blue thick dashed line is our model whereas the solid black line is $f(\mathcal{R})=\mathcal{R}$. %do not uncomment this% The $f(\mathcal{R})$ as a function of Ricci scalar $\mathcal{R}$, moving away from the center of the galaxy we have approximately linear relation between $f(\mathcal{R})$ and $\mathcal{R}$. 
	%		Moving slightly away from $r_{ini}$ our $f(\mathcal{R})$ model starts to deviate from linear relations very slightly (a clearer version of this statement is given in Table (\ref{table:rRf})). As we reach the boundary of the galaxy we have $\mathcal{R}\approxeq0$ and $f(\mathcal{R})\approxeq0$.}
	%	\label{fig:fR}
	%\end{figure}
	
	Figure (\ref{fig:f(R)}) is the parametric graph of $f(\mathcal{R})$ and $\mathcal{R}$. We see from the graphs that as we move away from the galactic center the line of $f(\mathcal{R})$ is approximately linear for vacuum.  For the other models with a significant deviation from the GR has been seen.%The solid black line is connecting (1,1) and (0,0) which %corresponds to $f(\mathcal{R})=\mathcal{R}$, it overlaps our model. The black line has been drawn to compare Einstein's gravity and our $f(\mathcal{R})$ gravity model. Due to very slight modification from linear  relation it is very hard to see any difference between $f(\mathcal{R})=\mathcal{R}$ and our model. But we also give some numerical values to see that difference.
	
	As mentioned before, the deviation of the $f(\mathcal{R})$ gravity from the GR for Fig. (\ref{fig:f(R)-sub2}) is permanent at the galactic edge which essentially implies that at the galactic edge we do not have 
	$$f(\mathcal{R})_{\text{inside solution}}\neq f(\mathcal{R})_{\text{outside the galaxy}}~,$$
	where $f(\mathcal{R})_{\text{outside the galaxy}}=\mathcal{R}$,\footnote[1]{If we consider that GR can explain all problems e.g. the dark energy etc.} if we try to explain the rotational velocity curves with the power law density distribution. Nevertheless, upon demanding the resolution to the rotational velocity curves problem in power law matter distribution, we must solve some other problem in $f(\mathcal{R})$ gravity, for example dark energy, and get 
	$$f(\mathcal{R})_{\text{inside solution}}= f(\mathcal{R})_{\text{outside the galaxy}}~,$$
	from the free parameters. This is the only way to rotational velocity curves' problems' resolution in $f(\mathcal{R})$ gravity. This behavior does not occur in the simple model of the galaxy with a core possibly because the core effect cancels out the occurrence of the permanent deviation from GR which does not happen in power law distribution.
	
	After plotting $f(\mathcal{R})$ from the numerical results we observed that graphically we cannot differentiate between obtained $f(\mathcal{R})$ and $f(\mathcal{R})=\mathcal{R}$ for vacuum. For this in-differentiability reason, we give some numerical values of $\mathnormal{f}$ and $\mathnormal{R}$ for all density profiles to see the difference between obtained $f(\mathcal{R})$ and Einstein's gravity in Table (\ref{table:rRf}).
	{\begin{center}
			\begin{sidewaystable}[htb!]		
				\bigskip
				\centering\tiny\setlength\tabcolsep{2pt}
				\vspace*{0cm}
				\hspace*{0cm}
				\begin{center}
					\begin{tabular}	
						{|c|c|c|c|c|c|c|c|c|c|c|c|c|}
						\hline
						Distance & \multicolumn{2}{|c|}{Vacuum solution} &  \multicolumn{2}{|c|}{$\rho(r)=\dfrac{1.073\times 10^{20}}{r^2}\text{Kg/m}$} & \multicolumn{2}{|c|}{$\rho(r)=\dfrac{1.05792\times10^{45}}{r^2}\text{Kg/m}$} & \multicolumn{2}{|c|}{$\rho(r)=\dfrac{1}{4\pi G} ~ \dfrac{V_t^2}{r^2+a_H^2}$} & \multicolumn{2}{|c|}{NFW profile - M31} & \multicolumn{2}{|c|}{NFW profile - Milky Way} \\ \hline
						$\space\space\space$ r (Kpc)\text{\space} & \text{\space} $\mathcal{R}_N$\text{\space} & \text{\space} $f_N$\text{\space} & \text{\space} $\mathcal{R}_N$\text{\space} & \text{\space} $f_N$\text{\space} & \text{\space} $\mathcal{R}_N$\text{\space} & \text{\space} $f_N$\text{\space} & \text{\space} $\mathcal{R}_N$\text{\space} & \text{\space} $f_N$\text{\space} & \text{\space} $\mathcal{R}_N$\text{\space} & \text{\space} $f_N$\text{\space} & \text{\space} $\mathcal{R}_N$\text{\space} & \text{\space} $f_N$\text{\space} \\ \hline					
						1       & 1              &  1            &  1          &  1				&  1               &  1           		& 1			 	  & 1	     		& 1			    & 1				& 1				& 1				\\ \hline
						2       & 0.249931       &  0.249868     &  0.182652   &  0.101821		&  0.453513        &  0.456658	 		& 0.2632          & 0.540145	    & 0.335823      & 0.694513		& 0.698097		& 1.44813		\\ \hline			
						3       & 0.111028       &  0.110944	 &  0.0315371  & -0.0639862		&  0.235097        &  0.266149			& 0.120486        & 0.376388		& 0.173885      & 0.550969		& 0.496062		& 1.59709		\\ \hline
						4       & 0.0624127      &  0.0623267  	 & -0.0213652  & -0.122151		&  0.139229        &  0.175047			& 0.0691891       & 0.291178		& 0.107793      & 0.462936		& 0.362425		& 1.60594		\\ \hline
						5       & 0.0399118      &  0.0398249  	 & -0.0467207  & -0.15044		&  0.0902852       &  0.124076			& 0.04499         & 0.238557		& 0.0738413     & 0.4019		& 0.272028		& 1.55168		\\ \hline
						6       & 0.0276907      &  0.0276042  	 & -0.0596686  & -0.164586 	    &  0.0624259       &  0.0925678			& 0.0316479       & 0.202675		& 0.053923      & 0.35645		& 0.209213		& 1.47148		\\ \hline
						7       & 0.0203219      &  0.0202357  	 & -0.0677877  & -0.173591		&  0.0452639       &  0.0716907			& 0.0235048       & 0.176564		& 0.041175      & 0.320982		& 0.164411		& 1.3833		\\ \hline
						8       & 0.0155396      &  0.0154544  	 & -0.0730347  & -0.179425		&  0.0340426       &  0.0571301	 		& 0.0181643       & 0.15667		    & 0.032497      & 0.292368		& 0.131682		& 1.29574		\\ \hline
						9       & 0.0122613      &  0.012178	 & -0.0765907  & -0.183349		&  0.0263568       &  0.0465658			& 0.0144702       & 0.140984		& 0.0263109     & 0.268703		& 0.107251		& 1.2127		\\ \hline
						10      & 0.00991653     &  0.00983378 	 & -0.079216   & -0.186187		&  0.020893        &  0.0386564 		& 0.0118071       & 0.128282		& 0.0217403     & 0.248751		& 0.0886558		& 1.13576		\\ \hline
						12      & 0.00686141     &  0.00677846 	 & -0.082374   & -0.18965		&  0.0138786       &  0.0278139 		& 0.00830326      & 0.10894		    & 0.0155597     & 0.216854		& 0.0629228		& 1.0012		\\ \hline
						15      & 0.00436394     &  0.00428967   & -0.0856186  & -0.193066		&  0.00831973      &  0.0183743 		& 0.00539486      & 0.0891646	    & 0.0102585     & 0.182163		& 0.0405983		& 0.842174		\\ \hline	
						20      & 0.00241698     &  0.00233415   & -0.0874745  & -0.195422		&  0.00423402      &  0.0105742 		& 0.00309169      & 0.0688688		& 0.0059315     & 0.14406		& 0.0226156		& 0.658383		\\ \hline
						40      & 0.000540227    &  0.000455242	 & -0.0895908  & -0.197809		&  0.000785432     &  0.00258876		& 0.000810811     & 0.0369127		& 0.00152798    & 0.0787609		& 0.00536312	& 0.342405		\\ \hline			
						50      & 0.000314892    &  0.000229789	 & -0.0898401  & -0.198078		&  0.000450777     &  0.00161228		& 0.000526248     & 0.0301774		& 0.000981485   & 0.0643019		& 0.00338257	& 0.275025		\\ \hline
						70      & 0.000118108    &  0.0000336948 & -0.0900324  & -0.198255		&  0.000193579     &  0.000776636		& 0.000274483     & 0.0222764		& 0.000502864   & 0.0471345		& 0.00169752	& 0.197045		\\ \hline			
						100     & 0.0000142981   & -0.0000707476 & -0.0902719  & -0.198397		&  0.0000784606    &  0.000351335		& 0.0001376       & 0.0161379		& 0.000247666   & 0.0337783		& 0.00082261	& 0.138115		\\ \hline			
						200     & -0.0000602979  & -0.000145592	 & -0.0904309  & -0.198332		&  0.0000134811    &  0.0000719514		& 0.0000359503    & 0.00861824		& 0.0000628803  & 0.0175837		& 0.00020366	& 0.0691108		\\ \hline			
						500     & -0.0000812865  & -0.000166571	 & -0.0915381  & -0.197183		&  $1.32106\times10^{-6}$ &  $8.12161\times10^{-6}$	& $6.08423\times10^{-6}$ & 0.00375134		& 0.0000103521  & 0.00739263	& 0.0000324832	& 0.0276519		\\ \hline		
						%1000   & -0.0000842671  & -0.000169521	 & -0.0957065  & -0.192657		&  2.12638*10^-7   &  1.22194*10^-6 	& 1.58575*10^-6   & 0.00199657		& 2.65023*10^-6 & 0.00383502	& 8.11739*10^-6	& 0.0138279		\\ \hline
						%2000   & -0.0000850294  & -0.000170253	 & -0.121262   & -0.166398		&  1.27299*10^-8   & -1.32758*10^-7 	& 4.13015*10^-7   & 0.00106134		& 6.78686*10^-7 & 0.00198855	& 2.02927*10^-6	& 0.00691491	\\ \hline
						%5000   & -0.0000853412  & -0.000170361	 & -43.9501    & -38.7925		& -2.86413*10^-8   & -3.72834*10^-7 	& 6.96585*10^-8   & 0.000459424		& 1.12075*10^-7 & 0.000834073	& 3.2471*10^-7	& 0.00276647	\\ \hline
						%10000  & -0.0000857383  & -0.000170023	 & -1111.91    & -1089.14		&  _ 		 	   &  _ 				& 1.81025*10^-8   & 0.000243504		& 2.86887*10^-8 & 0.00133365	& 8.11844*10^-8	& 28.0955		\\ \hline
						%100000 &  _			 &  _			 &  _		   &  _				&  _			   &  _					& 2.04931*10^-10  & 0.0000293251 	& _             & _ 			& _				& _				\\ \hline
					\end{tabular}
				\end{center}
				\caption{%Comparison of GR and our model. 
					Some numerical values of $\mathcal{R}$ and $f(\mathcal{R})$ for different galactic matter density profiles. The model starts from $f(\mathcal{R})=\mathcal{R}$, moving away from galactic center it deviates very slightly from GR.}
				\label{table:rRf}	 
				\hspace*{-1cm}
			\end{sidewaystable}
	\end{center}}
	%\newpage
	From the Table (\ref{table:rRf}), we observe that in vacuum a slight deviation from GR can explain the constant rotational velocity of a particle in moving around galactic center. For the vacuum solution specifically, a slight negative correction to $f(\mathcal{R})=\mathcal{R}$ can explain the observed flat rotation curves i.e. $f(\mathcal{R})<\mathcal{R}$ till $r\approx200\text{Kpc}$. For $r>200\text{Kpc}$, the correction is positive i.e. $f(\mathcal{R})>\mathcal{R}$. %Whereas, for other cases in Table (\ref{table:rRf}), $\mathcal{R}_N$ and $f_N$ change their relation from $|f_N|>|\mathcal{R}_N|$ to $|f_N|<|\mathcal{R}_N|$ around $35.3\text{Kpc}$ % as suggested by \citet{refId0-1,refId0-2}.
	
	For all the density profiles considered in this paper, a significant difference from GR is observed because of the added mass in the galaxy. The matter density profile \ref{eq:Power-law-density-profile} demands the resolution of some other problem in the $f(\mathcal{R})$ gravity such that matching $f(\mathcal{R})$ gravity outside and inside the galactic environment. This profile essentially assumes that the tangential velocity of a particle moving around the galaxy is constant which we took to be $300\text{Km/s}$ for Figs. (\ref{fig:Lambda-sub2}), (\ref{fig:Ricci(r)-sub2}), (\ref{fig:fr-sub2}) and (\ref{fig:f(R)-sub2}) %all (b) graphs in the figures 
	and $1.7\times 10^8 \text{Km/s}$ for (\ref{fig:Lambda-sub6}), (\ref{fig:Ricci(r)-sub6}), (\ref{fig:fr-sub6}) and (\ref{fig:f(R)-sub6}).%all (f) graphs in the figures. 
	The second case has been taken to see the extremal deviation of modified gravity from GR since in the second case we take $V_t\longrightarrow c$, where $c$ is speed of light.
	
	In the calculations, the precision of the numerical values is the machine precision of the MATHEMATICA software set at $15.9546$ whereas the accuracy goal is set to $10$. Thus, we expect the numerical error in each result to be less than or equal to $10^{-10}+|x|10^{-15.9546/2}$, where $|x|$ is $\lambda$, $\mathcal{R}$ and $f(\mathcal{R})$ at different values of $r$.
	%%%%%%%%%%%%%%%%%%%%%%%%%%%%%%%%%%%%%%%%%%%%%%%%%%%%%%%%%%%%%%%%%%%%%%%%%%%%%%%%%%%%%%%%
	\section[Scalar dark matter field from $f(\mathcal{R})$ gravity]{Scalar dark matter field from $f(\mathcal{R})$ gravity}
	%%%%%%%%%%%%%%%%%%%%%%%%%%%%%%%%%%%%%%%%%%%%%%%%%%%%%%%%%%%%%%%%%%%%%%%%%%%%%%%%%%%%%%%%
	The $f(\mathcal{R})$ gravity action in vacuum (as described in the previous chapter) can be written as
	\begin{equation}
	S=\frac{1}{2\kappa^{2}}\int \mathrm{d}^{4}x\sqrt{-g}\,
	\left[ f(\chi)+f_{,\chi}(\chi) (\mathcal{R}-\chi) \right]\,,
	\label{BifR--}
	\end{equation}
	with the real scalar field $\chi=\mathcal{R}$ constrained with $f,_{\chi\chi}\neq0$. Using conformal transformation of the metric as $\tilde{g}_{\mu\nu}=F g_{\mu\nu}$, one can transform the above action to an action of a real scalar field $\phi$ where
	\begin{equation}\label{phi}
	\phi=\sqrt{\dfrac{3}{2}}\dfrac{1}{\kappa}\log(F)~,
	\end{equation}
	which is minimally coupled with the Einstein gravity with the potential
	\begin{equation}
	V(\varphi)=\frac{\mathcal{R}(\varphi)\,F(\mathcal{R}(\varphi))-f(\mathcal{R}(\varphi))}
	{2\kappa^2 F(\mathcal{R}(\varphi))^2}\,.
	\label{Uphidef--}
	\end{equation}
	Since we have obtained few modified gravity functions for different galactic scenarios in the previous sections, we can thus construct their scalar Brans-Dicks analogue field and therefore obtain some information about DM scalar field using modified gravity.
	%-------------------------------
	\subsection{Analytic vacuum solution I, of ref. \citet{Usman2016}} \label{VS-I}
\begin{figure}%[h!]
		\centering
		\begin{minipage}{0.48\linewidth}
			\centering
			{\includegraphics[height=0.8\textwidth]{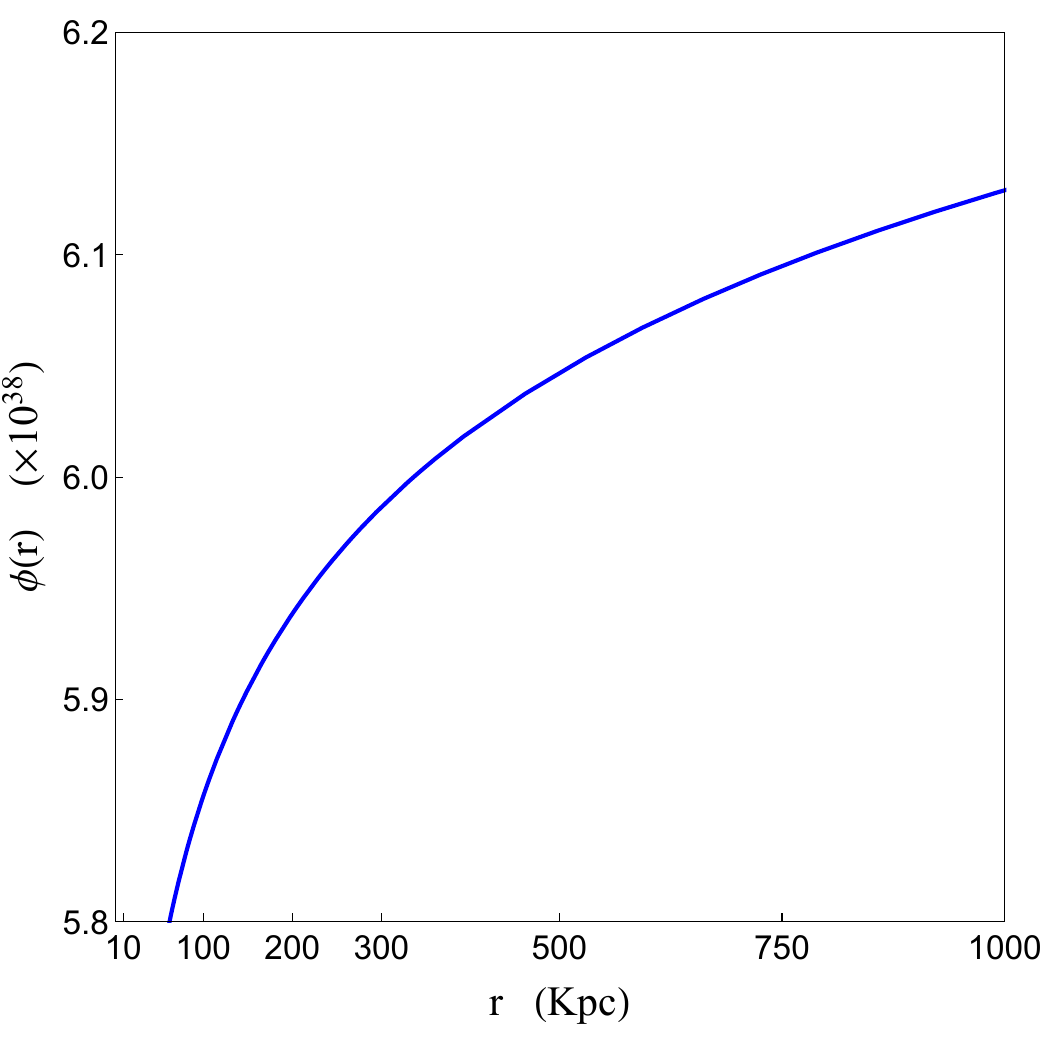}}
			\subcaption{The DM field in vacuum.}
			\label{fig:PhirVacuumAnalyticSoln1}
		\end{minipage}
		\hspace{0.1cm}
		\begin{minipage}{0.48\linewidth}
			\centering
			{\includegraphics[height=0.8\textwidth]{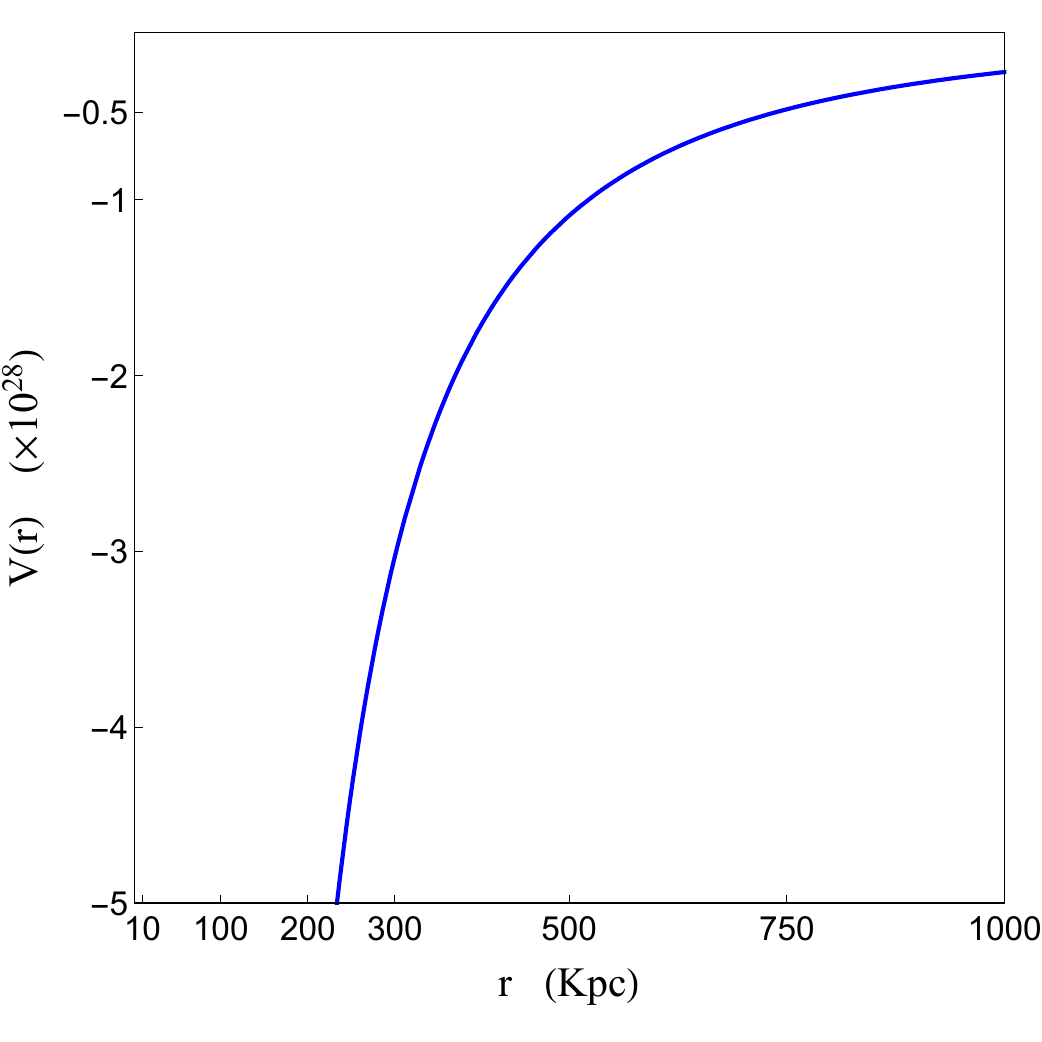}}
			\subcaption{The field potential.}
			\label{fig:VrVacuumAnalyticSoln1}
		\end{minipage}
		%\\[2ex]
		%	\begin{minipage}{0.95\linewidth}
		%		\centering
		%		\includegraphics[height=0.6\textwidth]{Plots/Scalar-DM/VPhiVacuumAnalyticSoln1.pdf}
		%		\caption{The field potential as a function of field.}
		%		\label{fig:VPhiVacuumAnalyticSoln1}
		%	\end{minipage}
		\caption{The scalar field and potential properties in vacuum as a function of radial distance from galactic center. %for Sec. (\ref{Case-I}) solution.
		}
		\label{fig:Phi-V-AnalyticSoln1a}
	\end{figure}

\begin{figure}%[h!]
		\centering
		%\\[2ex]
		\begin{minipage}{0.95\linewidth}
			\centering
			\includegraphics[height=0.4\textwidth]{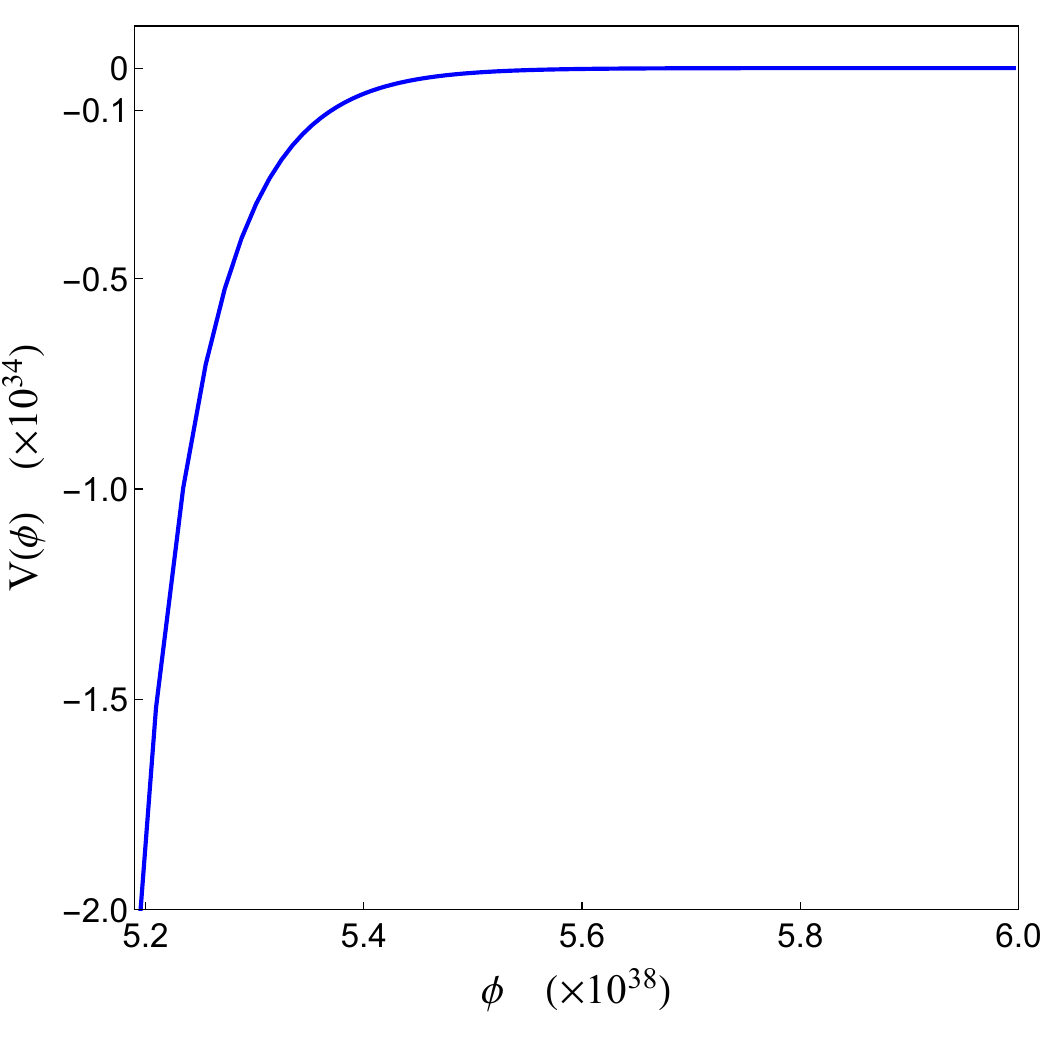}
			%\caption{}
			\label{fig:VPhiVacuumAnalyticSoln1b}
		\end{minipage}
		\caption{The field potential as a function of field. %for Sec. (\ref{Case-I}) solution.
		}
		\label{fig:Phi-V-AnalyticSoln1b}
	\end{figure}

	Using the results of \citet{Usman2016} into Eqs. (\ref{phi}), (\ref{Uphidef--}), we obtain 
	\begin{equation}\label{phi-soln-1--Case-I}
	\phi(r)=\sqrt{\dfrac{3}{2}} \dfrac{1}{\kappa} \ln \left(F_0 r^{2 m}\right)~,
	\end{equation}
	and
	\begin{equation}\label{V-soln-1--Case-I}
	V(r)=-\dfrac{6m^2}{54 \kappa^2 (1 - m)^2 F_0 }~r^{-2(1+m)}~.
	\end{equation}
	With $\kappa=8\pi G/c^4~, m=10^{-6} \text{ and }F_0=1$, the plots for field $\phi$, the potential $V$ and their parametric form are given by Figs. (\ref{fig:Phi-V-AnalyticSoln1a}) and (\ref{fig:Phi-V-AnalyticSoln1b})
	
	The graphs tell us that the field increases as does the potential upon moving away from the galactic center thus less and less DM upon moving away from galactic center. The potential also increases as a function of field value, and attains almost constant value. For small values of field the potential acts as the attractor. Thus, the field's energy density would decrease in moving away from the galactic center. This trend is similar to the Newtonian trend with different functional dependence.
	
	%-------------------------------
	%-------------------------------
	%-------------------------------
	\subsection{Analytic vacuum solution II, of ref. \citet{Usman2016}}\label{VS-II}
\begin{figure}%[h!]
		\centering
		\begin{minipage}{0.48\linewidth}
			\centering
			{\includegraphics[height=0.8\textwidth]{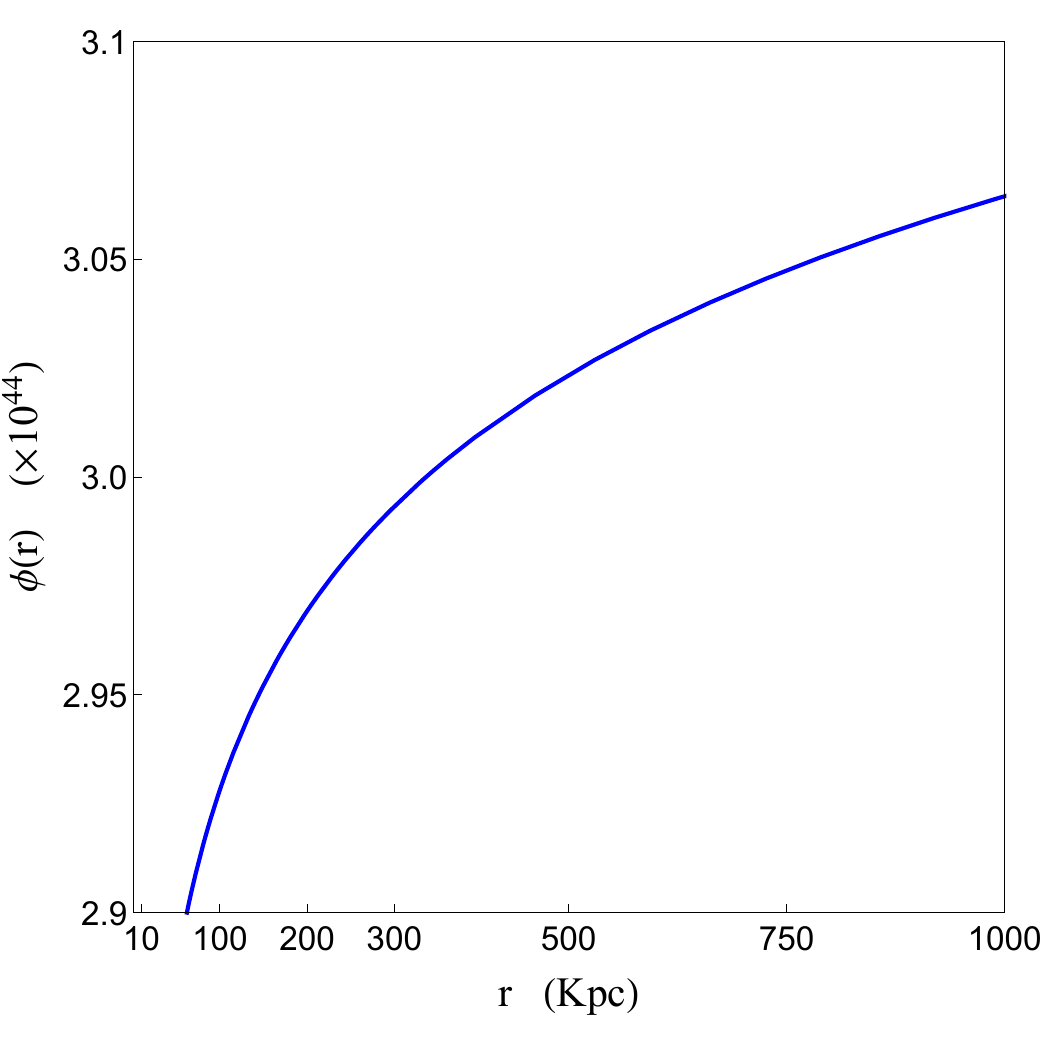}}
			\subcaption{The DM field in vacuum.}
			\label{fig:PhirVacuumAnalyticSoln2}
		\end{minipage}
		\hspace{0.1cm}
		\begin{minipage}{0.48\linewidth}
			\centering
			{\includegraphics[height=0.8\textwidth]{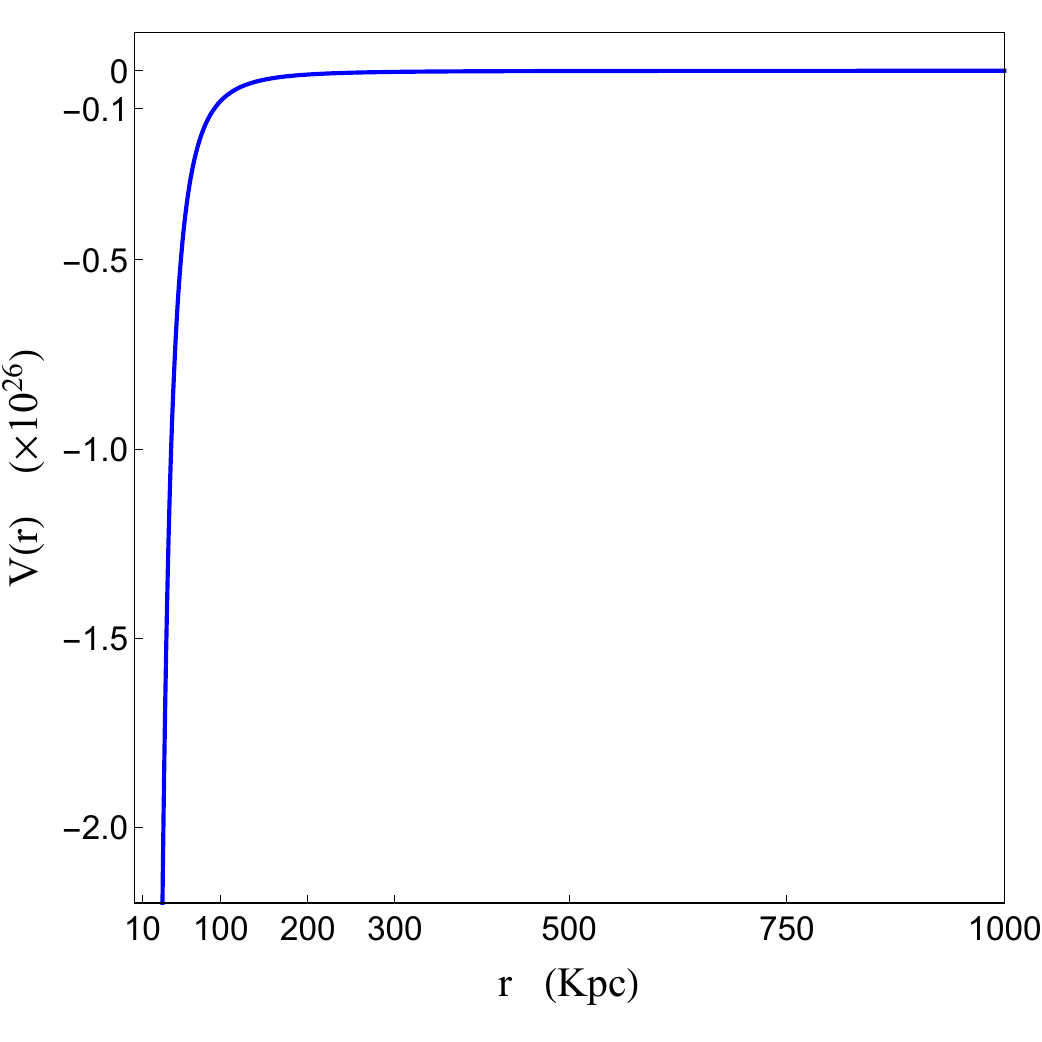}}
			\subcaption{The field potential.}
			\label{fig:VrVacuumAnalyticSoln2}
		\end{minipage}
		%\\[2ex]
		%	\begin{minipage}{0.95\linewidth}
		%		\centering
		%		\includegraphics[height=0.6\textwidth]{Plots/Scalar-DM/VPhiVacuumAnalyticSoln2.pdf}
		%		\caption{The field potential as a function of field.}
		%		\label{fig:VPhiVacuumAnalyticSoln2}
		%	\end{minipage}
		\caption{The scalar field and potential properties in vacuum as a function of radial distance from galactic center. %for Sec. (\ref{Case-II}) solution.
		}
		\label{fig:Phi-V-AnalyticSoln2a}
	\end{figure}

\begin{figure}%[h!]
		\centering
		%\\[2ex]
		\begin{minipage}{0.95\linewidth}
			\centering
			\includegraphics[height=0.4\textwidth]{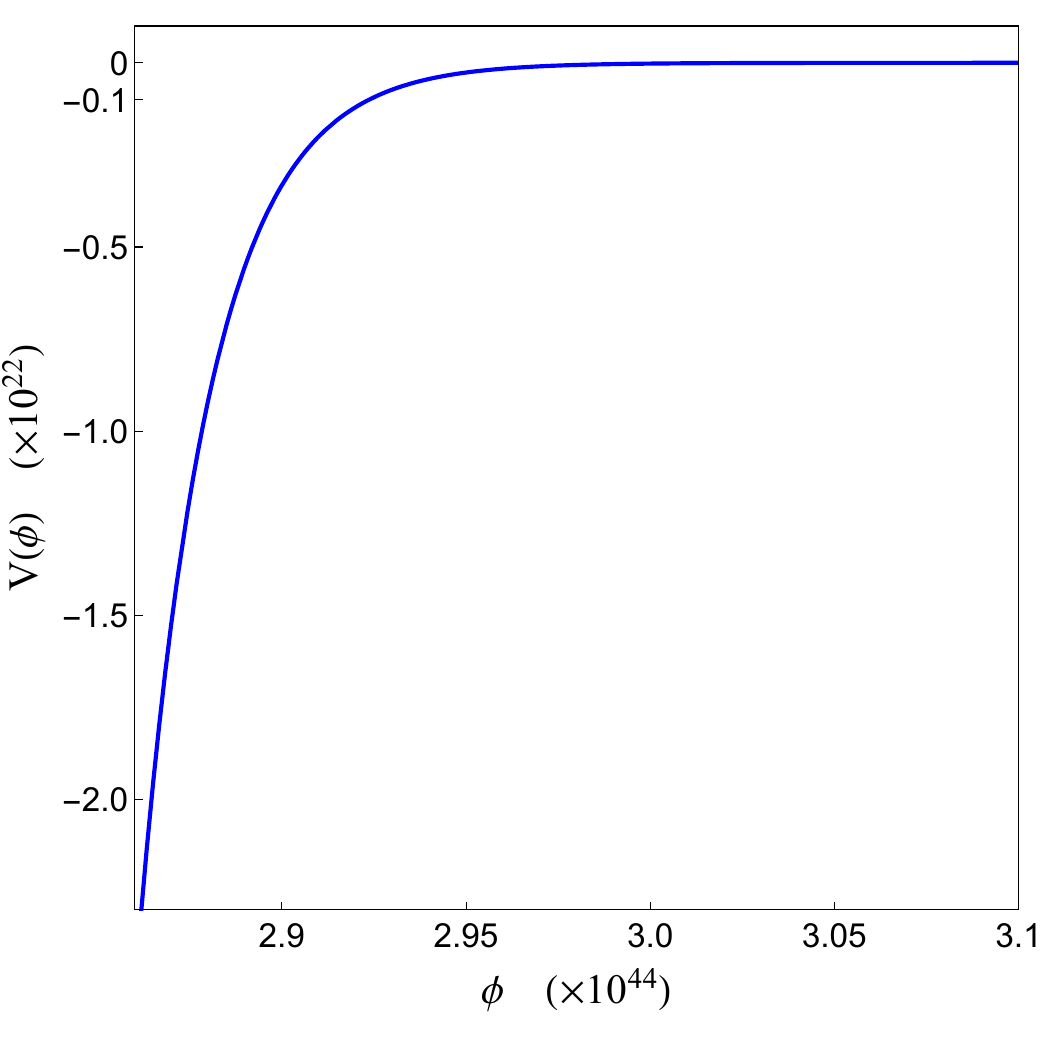}
			%\caption{}
			\label{fig:VPhiVacuumAnalyticSoln2}
		\end{minipage}
		\caption{The field potential as a function of field. %for Sec. (\ref{Case-II}) solution.
		}
		\label{fig:Phi-V-AnalyticSoln2b}
	\end{figure}

	Using the results of \citet{Usman2016} into Eqs. (\ref{phi}), (\ref{Uphidef--}), we obtain  
	\begin{equation}\label{phi-soln-1--Case-II}
	\phi(r)=\sqrt{\dfrac{3}{2}} \dfrac{1}{\kappa} \ln \left(F_0 r^{1-m}\right)~,
	\end{equation}
	and
	\begin{equation}\label{V-soln-1--Case-II}
	V(r)= \dfrac{(m-1)}{\kappa^2 F_0 (1 + m)^2} r^{-3+m}
	~.
	\end{equation}
	With $\kappa=8\pi G/c^4~, m=10^{-6} \text{ and }F_0=1$, the plots for field $\phi$, the potential $V$ and their parametric form are given by Figs. (\ref{fig:Phi-V-AnalyticSoln2a}) and (\ref{fig:Phi-V-AnalyticSoln2b})
	
	Just like analytic vacuum solution I, the above graphs tell that upon moving away from the galactic center the potential increases as does the field. The paramtric form of potential too increases with the increment in field value, which attains later constant value. For small field values the potential acts as the attractor. Thus, the field's density would decrease in moving away from the galactic center. This trend again is similar to the Newtonian trend with of course different functional dependence.
	%-------------------------------
	%-------------------------------
	%-------------------------------
	\subsection{Numerical power law density profile solution} \label{Numerical-Power-Law-Density-Profile-Solution}
\begin{figure}%[h!]
		\centering
		\begin{minipage}{0.48\linewidth}
			\centering
			{\includegraphics[height=0.8\textwidth]{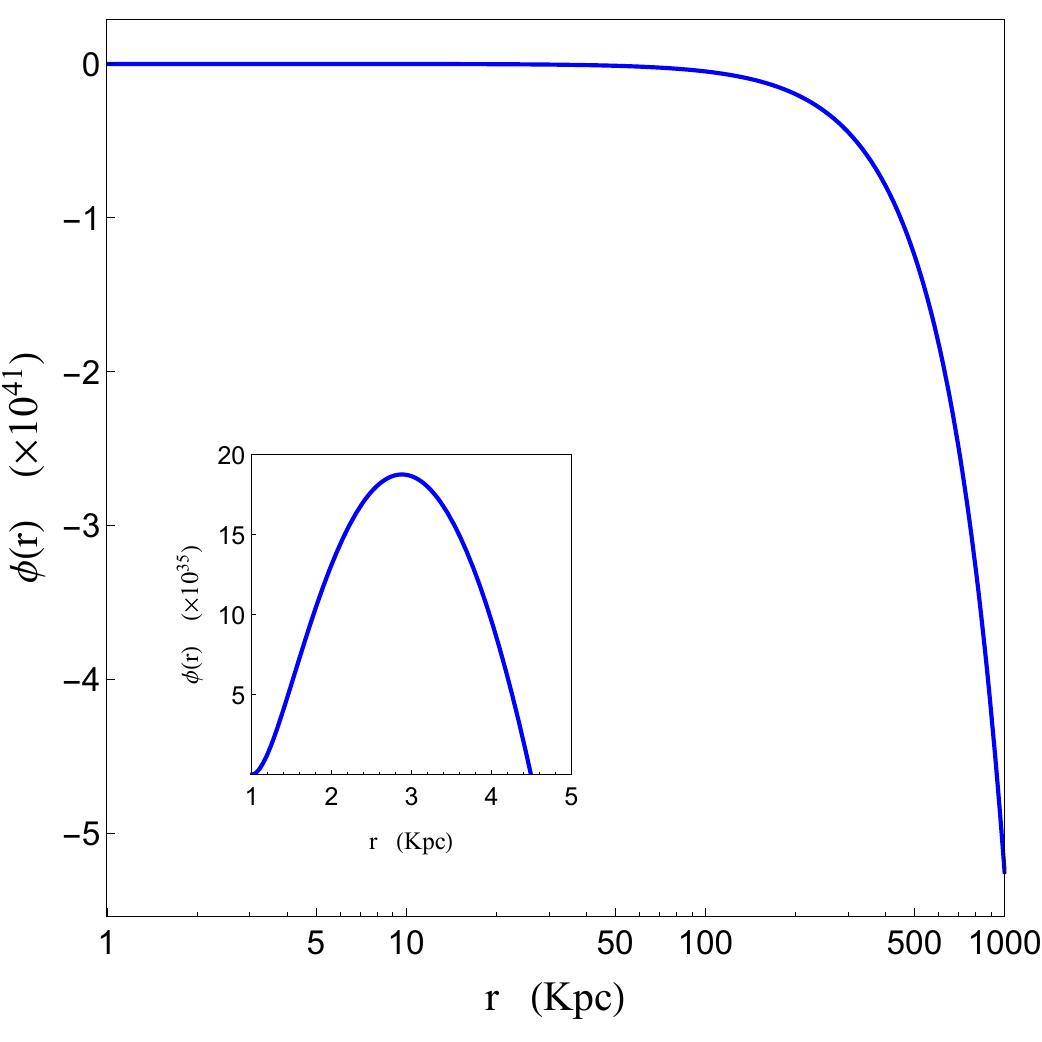}}
			\subcaption{The numerical solution of Eqs. (\ref{eq:1st}) and (\ref{eq:2nd}) for DM field.}
			\label{fig:phirpowerlawfunction}
		\end{minipage}
		\hspace{0.1cm}
		\begin{minipage}{0.48\linewidth}
			\centering
			{\includegraphics[height=0.8\textwidth]{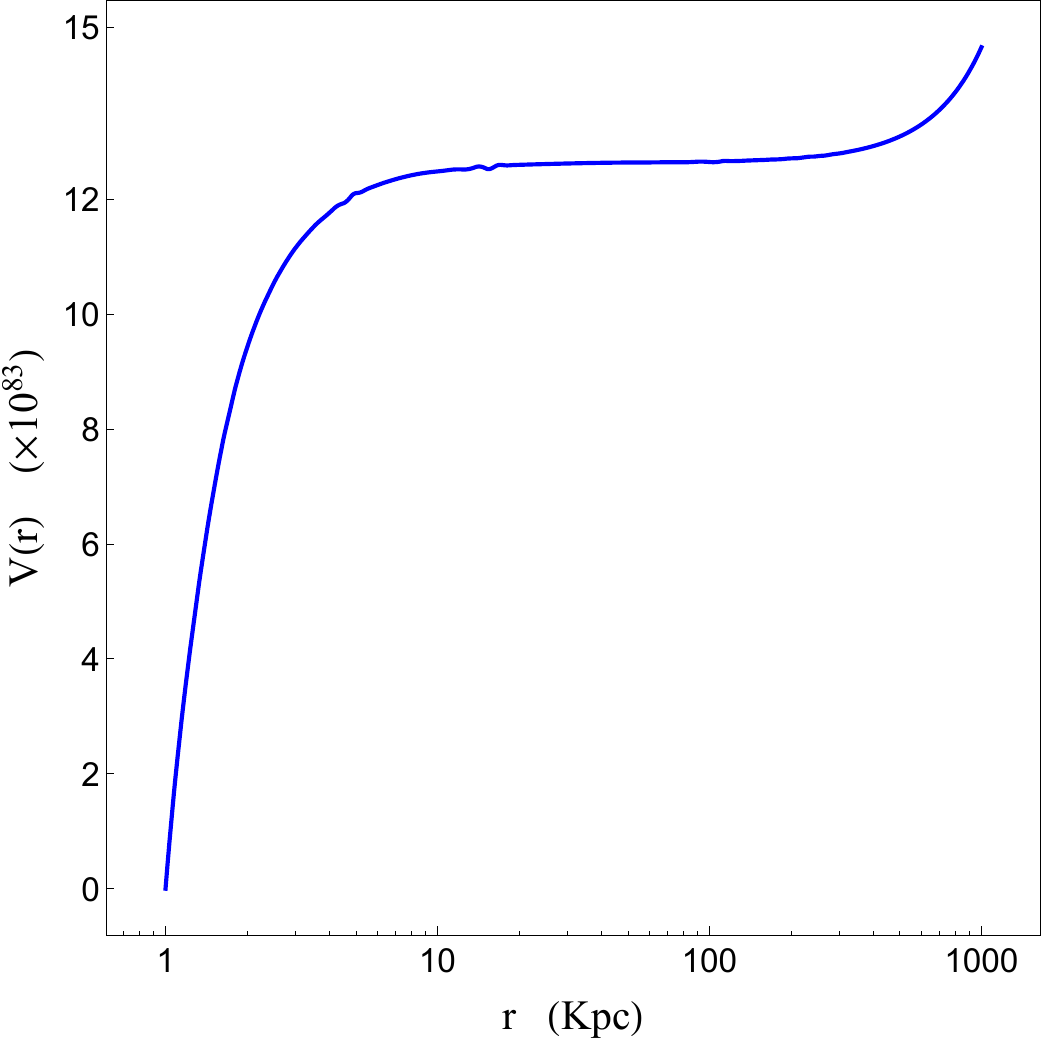}}
			\subcaption{The numerical solution of Eqs. (\ref{eq:1st}) and (\ref{eq:2nd}) for field potential.}
			\label{fig:Vrpowerlawfunction}
		\end{minipage}
		\caption{The scalar field properties for density profile given by Eq. (\ref{eq:Power-law-density-profile})}
		\label{fig:Phi-V-powerlawfunction-a}
	\end{figure}

\begin{figure}%[h!]
		\centering
		\includegraphics[height=0.4\textwidth]{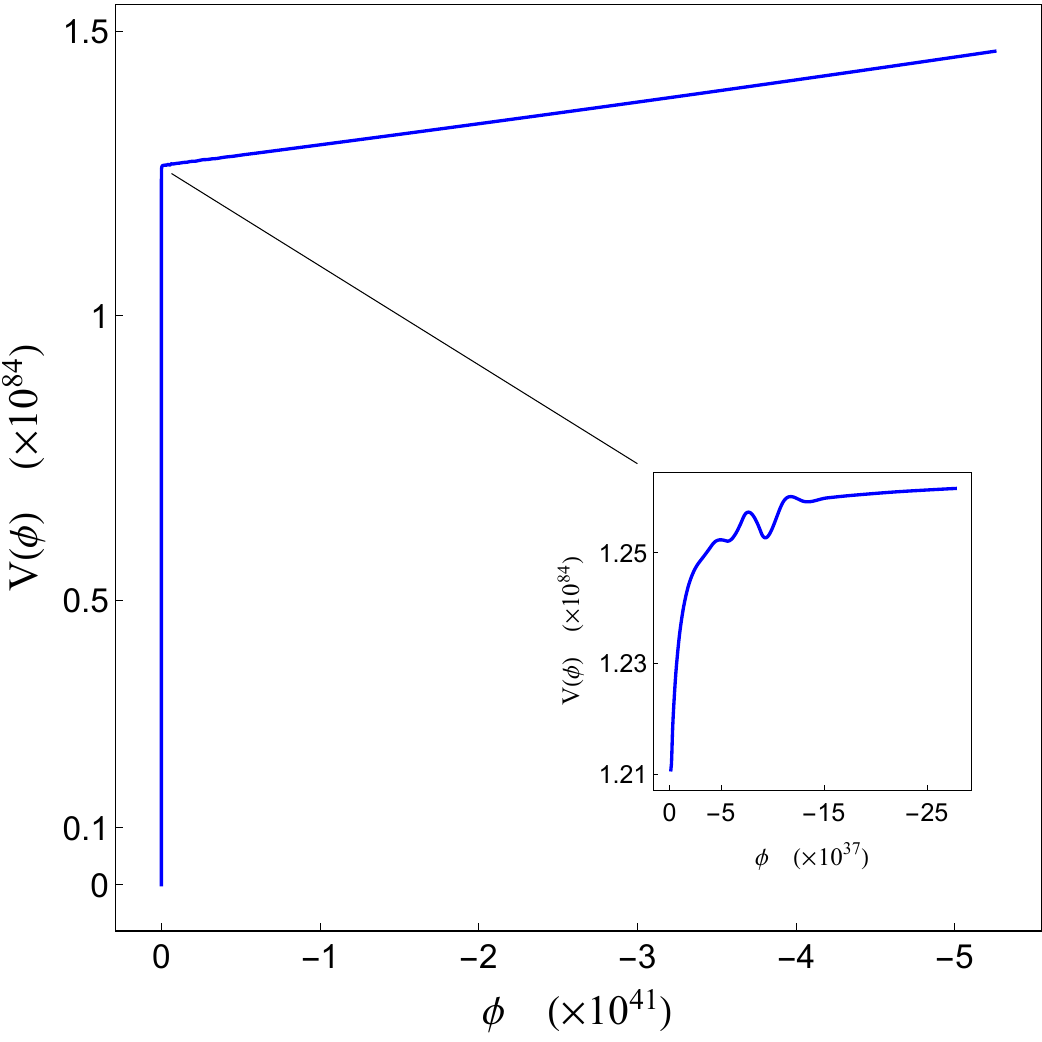}
		\caption{The parametric form from numerical solution of Eqs. (\ref{eq:1st}) and (\ref{eq:2nd}) for field potential for density profile given by Eq. (\ref{eq:Power-law-density-profile}).}
		\label{fig:Vphipowerlawfunction}
	\end{figure}

	Since it has been proven in the previous sections that we cannot obtain the analytic solution for all (and most importantly the desired) values of tangential speeds and possibly with any Newtonian matter's addition. Thus, we obtain the numerical solutions of Eqs. (\ref{eq:1st}) and (\ref{eq:2nd}). The obtained solutions are given in Figs. (\ref{fig:Phi-V-powerlawfunction-a}) and (\ref{fig:Vphipowerlawfunction})
	
	From the graphs, we see that in the increasing direction of radial distance from galactic center the field value first increases (shown by the small graphs in Fig. (\ref{fig:phirpowerlawfunction})) and then decreases. The potential on the other hand increases for up to approximately $10$Kpc then becomes constant until it starts increasing again at about $300$Kpc. The minimum value of the potential is zero which is achieved when the field also has zero value. The parametric form of potential tells us that as the field decreases, the potential increases thus showing an (approximate) inverse relation between field and potential. As the field value decreases moving away from the galactic center, this result shows that field as we move away from the galactic center the density of the dark matter would start to increase to compensate for the decreasing normal matter density. This is the only way that we can keep the tangential velocity constant and was expected. We must recall that we inserted a Newtonian matter with specific distribution in the field equations to obtain these numerical solution where the tangential speed around a galaxy does not depend upon the radial distance, the introduction of matter here perhaps also introduces interaction between the scalar DM field and normal matter which intern changes the dependence of potential.
	
	%-------------------------------
	%-------------------------------
	%-------------------------------
	\subsection{Numerical simple glactic model with a core solution}
	\label{Numerical-Simple-Glactic-Model-with-a-Core-Solution}
\begin{figure}%[h!]
		\centering
		\begin{minipage}{0.48\linewidth}
			\centering
			{\includegraphics[height=0.8\textwidth]{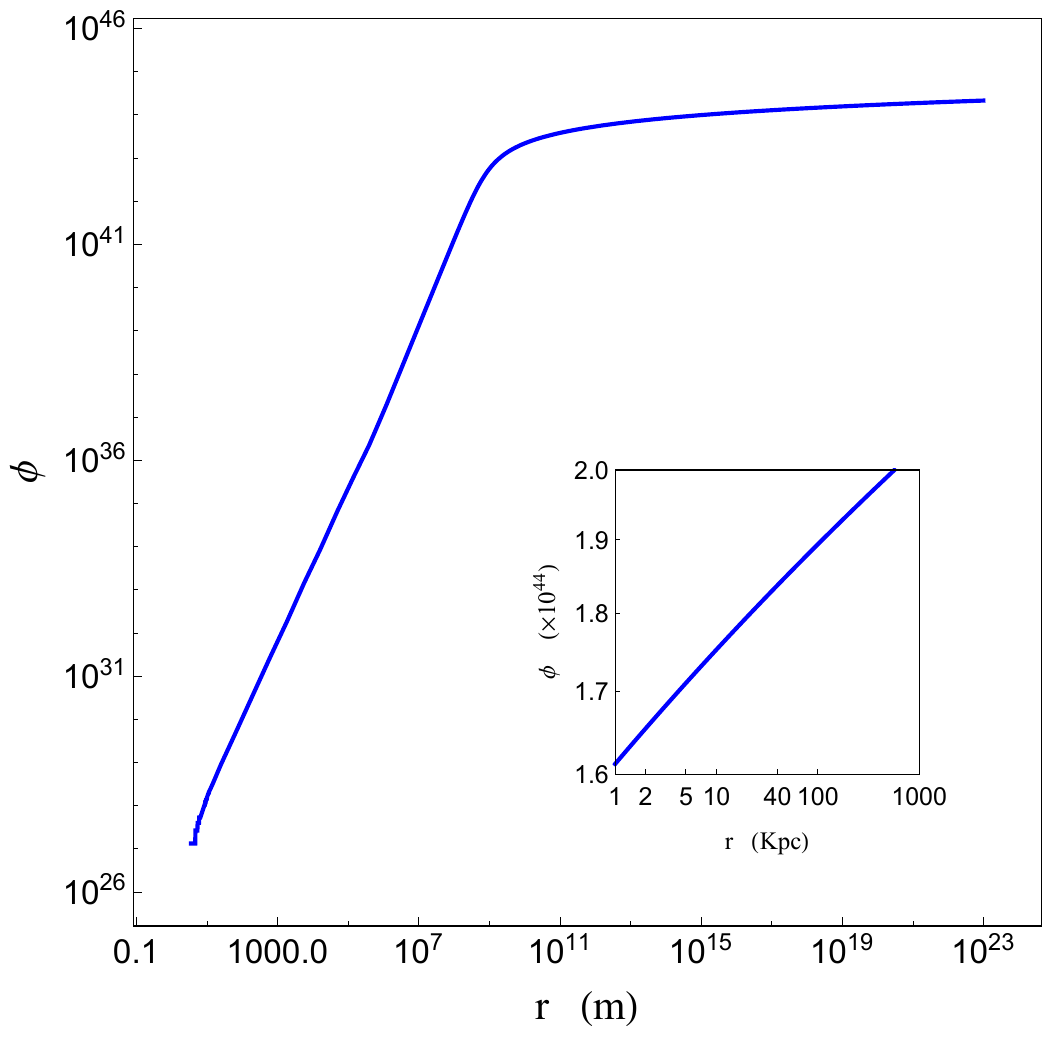}}
			\label{fig:phirSGM}
			\subcaption{The numerical solution of Eqs. (\ref{eq:1st}) and (\ref{eq:2nd}) for field potential.}
		\end{minipage}
		\hspace{0.1cm}
		\begin{minipage}{0.48\linewidth}
			\centering
			{\includegraphics[height=0.8\textwidth]{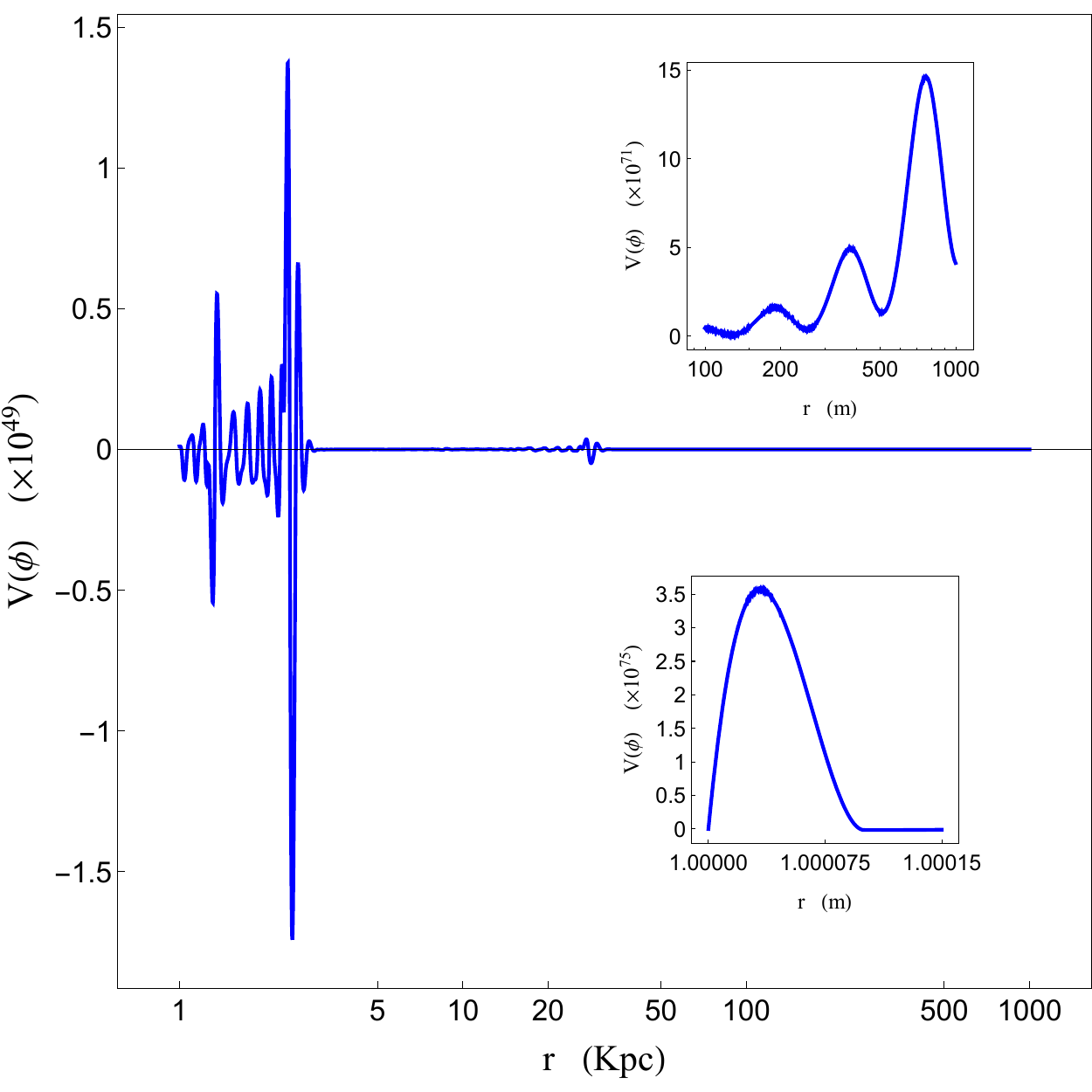}}
			\label{fig:VrSGM}
			\subcaption{The scalar field properties from numerical solution of Eqs. (\ref{eq:1st}) and (\ref{eq:2nd}) for field potential for density profile given by Eq. (\ref{eq:Simple-model-for-a-galaxy-with-a-core}).}
		\end{minipage}
		\caption{The scalar field properties from numerical solution of Eqs. (\ref{eq:1st}) and (\ref{eq:2nd}) for field potential for density profile given by Eq. (\ref{eq:Simple-model-for-a-galaxy-with-a-core}).}
		\label{fig:Phi-V-SGM}
	\end{figure}

\begin{figure}%[h!]
		\centering
		\includegraphics[height=0.4\textwidth]{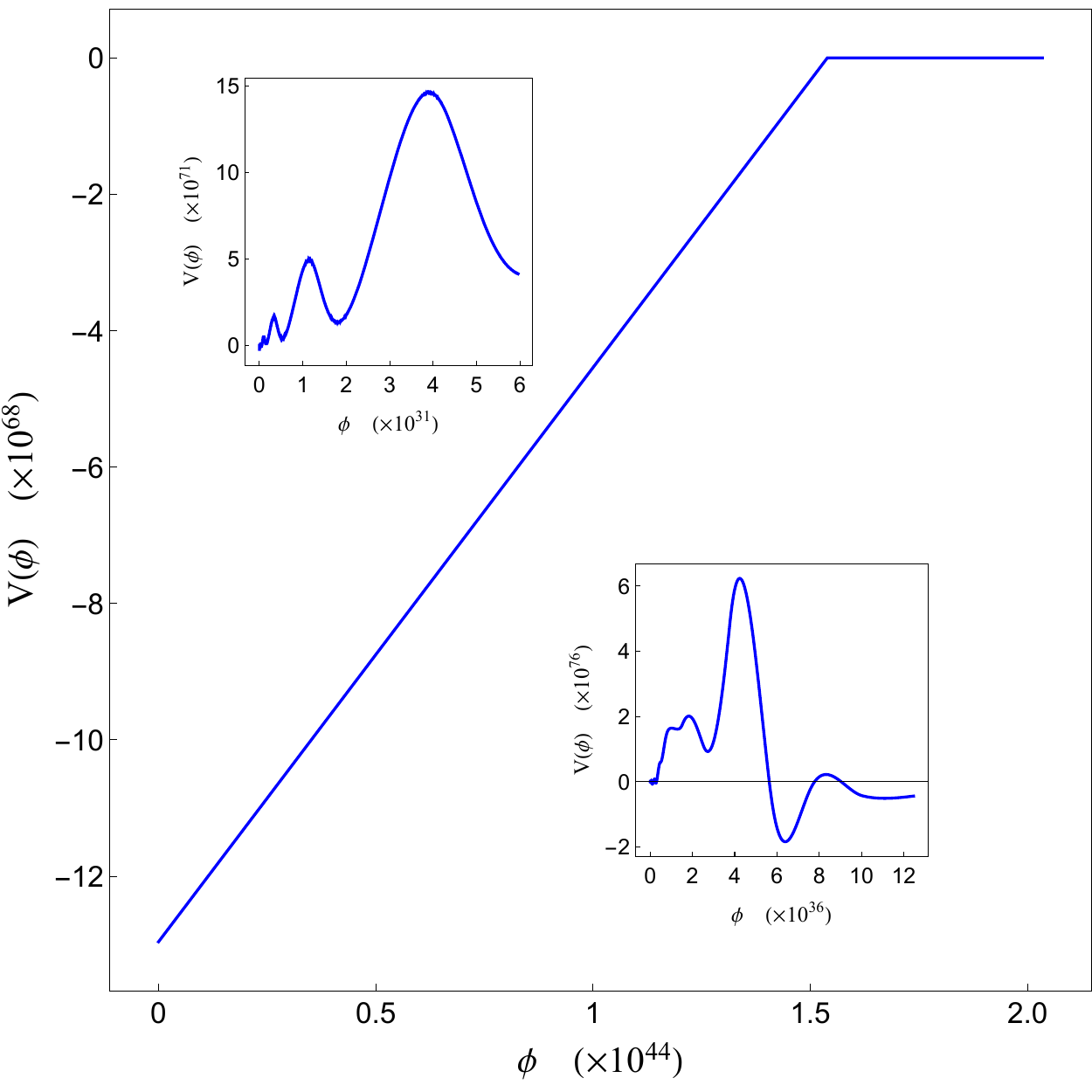}
		\caption{The parametric form from numerical solution of Eqs. (\ref{eq:1st}) and (\ref{eq:2nd}) for field potential for density profile given by Eq. (\ref{eq:Simple-model-for-a-galaxy-with-a-core}).}
		\label{fig:VphiSGM}
	\end{figure}

	On the footsteps of previous section, due to inability in getting the analytic solution,  we obtain the numerical solutions of Eqs. (\ref{eq:1st}) and (\ref{eq:2nd}). The obtained solutions are then plotted in Figs. (\ref{fig:Phi-V-SGM}) and (\ref{fig:VphiSGM})
	
	The graphs tell us that in the increasing direction of radial distance from galactic center the field value increases. The small boxed graph is for distance in Kpc. The field potential is quite different and interesting here, the upper small box is rather more interesting here, in it we see that the potential is increasing with small dips in it. Physically it tells that the DM field does not gravitate the same way as we anticipate in Newtonian case. The DM in this case has some preferred (more than one) collapsing rings\footnote{An analytical mathematical relation cannot be obtained here as we cannot solve the differential equations analytically (proved earlier). The multiple peaks in the Fig. (\ref{fig:Phi-V-SGM}) shows that the DM is collapsing (or is concentrated) at different radial locations}. We need to understand that we used a galactic environment in which the desired pattern of the tangential speed was achieved by the Newtonian matter. Thus, the small fraction of DM obtained here might parhaps belongs to the so called very famous \emph{Unparticle Physics} or some even unexplored non-talked Physics. We observe similar behavior of potential in the parametric form which is given by the small boxed graphs in Fig. (\ref{fig:VphiSGM}).
	
	We must also understand that the DM obtained in this scenario only contributes to a small factor in making the rotational velocity curves constant. This is because of the fact that the matter we introduced in EFEs already makes the rotational velocity curves as desired. The DM field could therefore because of its interaction with the introduced matter fluctuates the rotational speed from about $300$Km/h and we can get exact rotational velocity curves as obtained by Vera Rubin in \citet{vera1,vera2} If we know how the DM field interacts with the Newtonian matter at different $r$.
	%-------------------------------
	%-------------------------------
	%-------------------------------
	\subsection{Numerical NFW Milky way solution}\label{Numerical-NFW-Milky-Way-Solution}
\begin{figure}%[h!]
		\centering
		\begin{minipage}{0.48\linewidth}
			\centering
			{\includegraphics[height=0.8\textwidth]{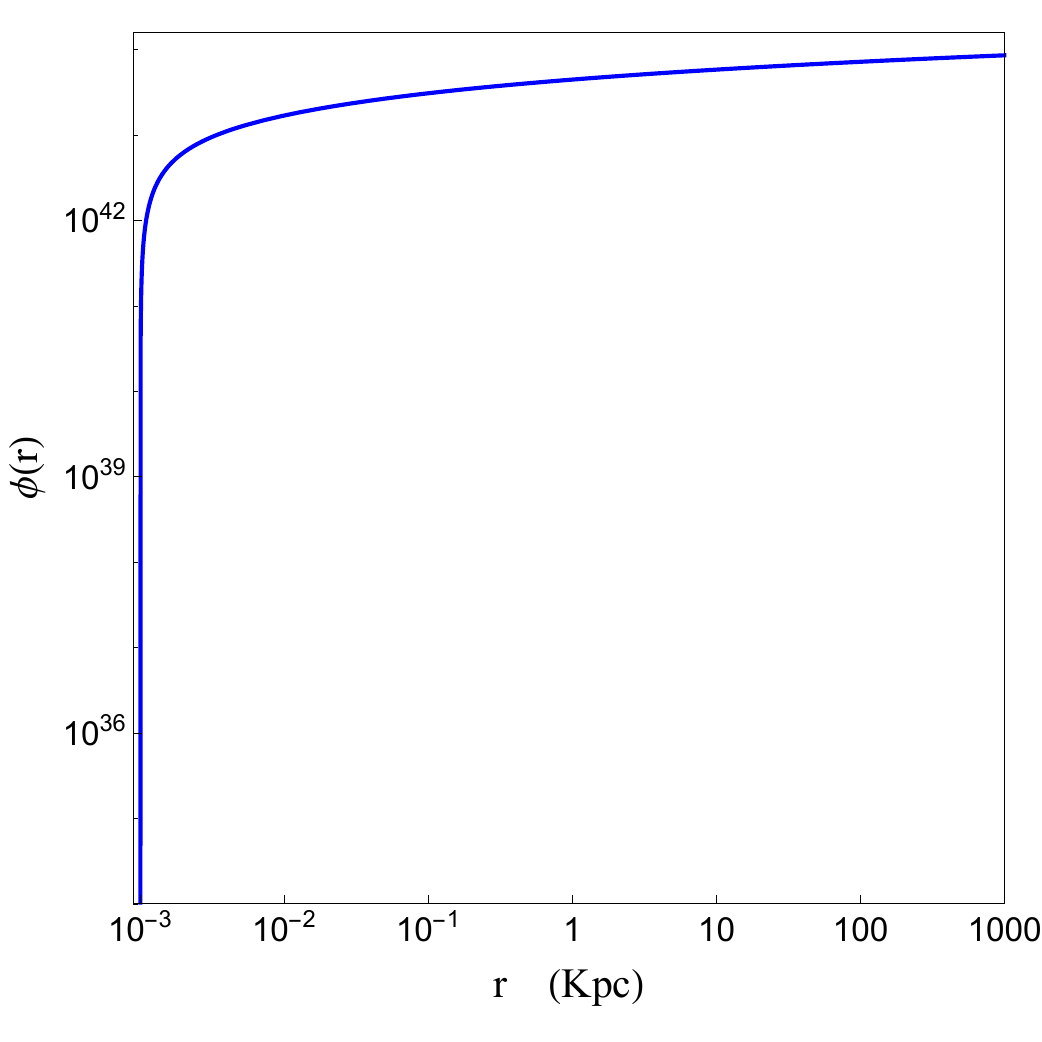}}
			\subcaption{The DM field in vacuum.}
			\label{fig:phirNFWMilkyWay}
		\end{minipage}
		\hspace{0.1cm}
		\begin{minipage}{0.48\linewidth}
			\centering
			{\includegraphics[height=0.8\textwidth]{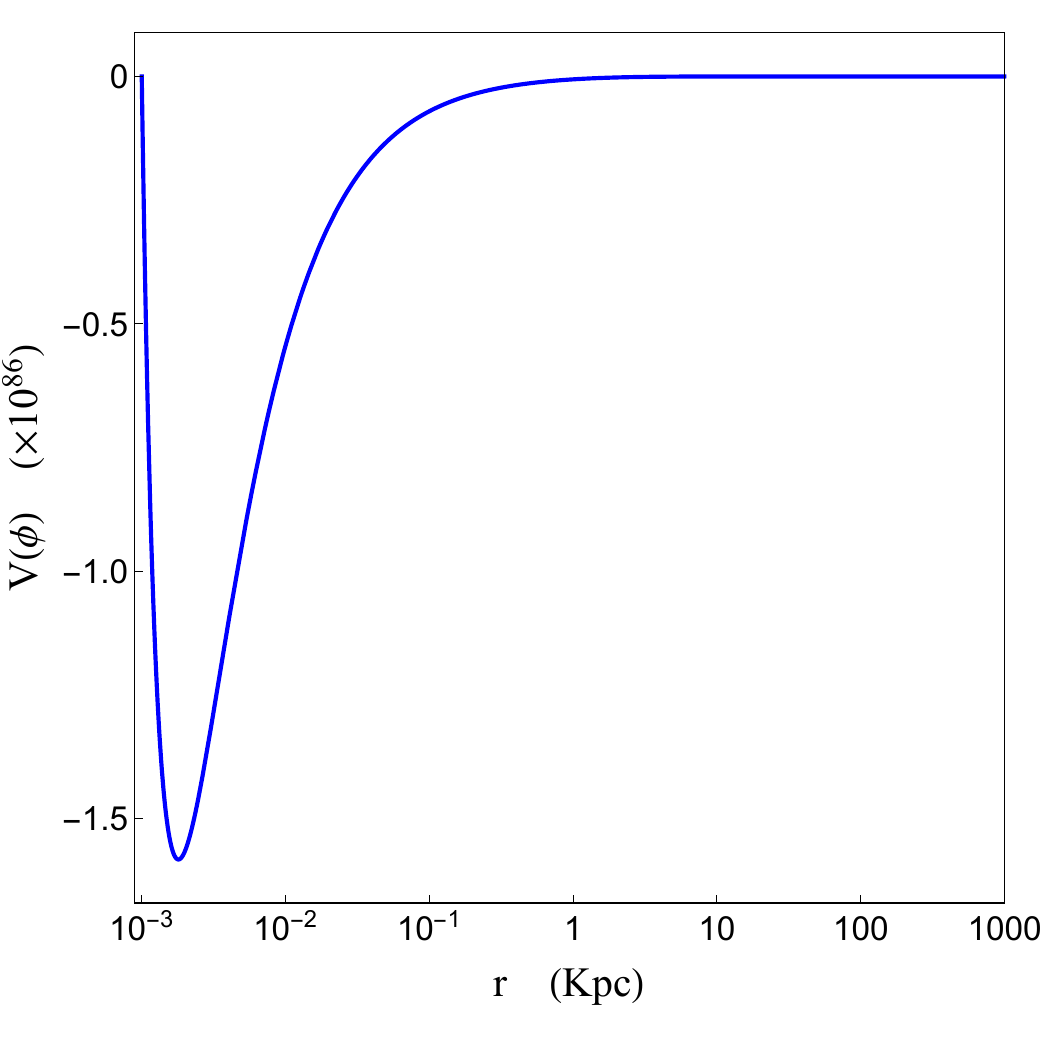}}
			\subcaption{The field potential.}
			\label{fig:VrNFWMilkyWay}
		\end{minipage}
		%\\[2ex]
		%	\begin{minipage}{0.95\linewidth}
		%		\centering
		%		\includegraphics[height=0.6\textwidth]{Plots/Scalar-DM/VphiNFWMilkyWay.pdf}
		%		\caption{The field potential as a function of field.}
		%		\label{fig:VphiNFWMilkyWay}
		%	\end{minipage}
		\caption{The scalar field properties in vacuum for NFW density profile with Milky way galaxy parameters}
		\label{fig:Phi-V-NFWMilkyWay}
	\end{figure}

\begin{figure}%[h!]
		\centering
		\includegraphics[height=0.4\textwidth]{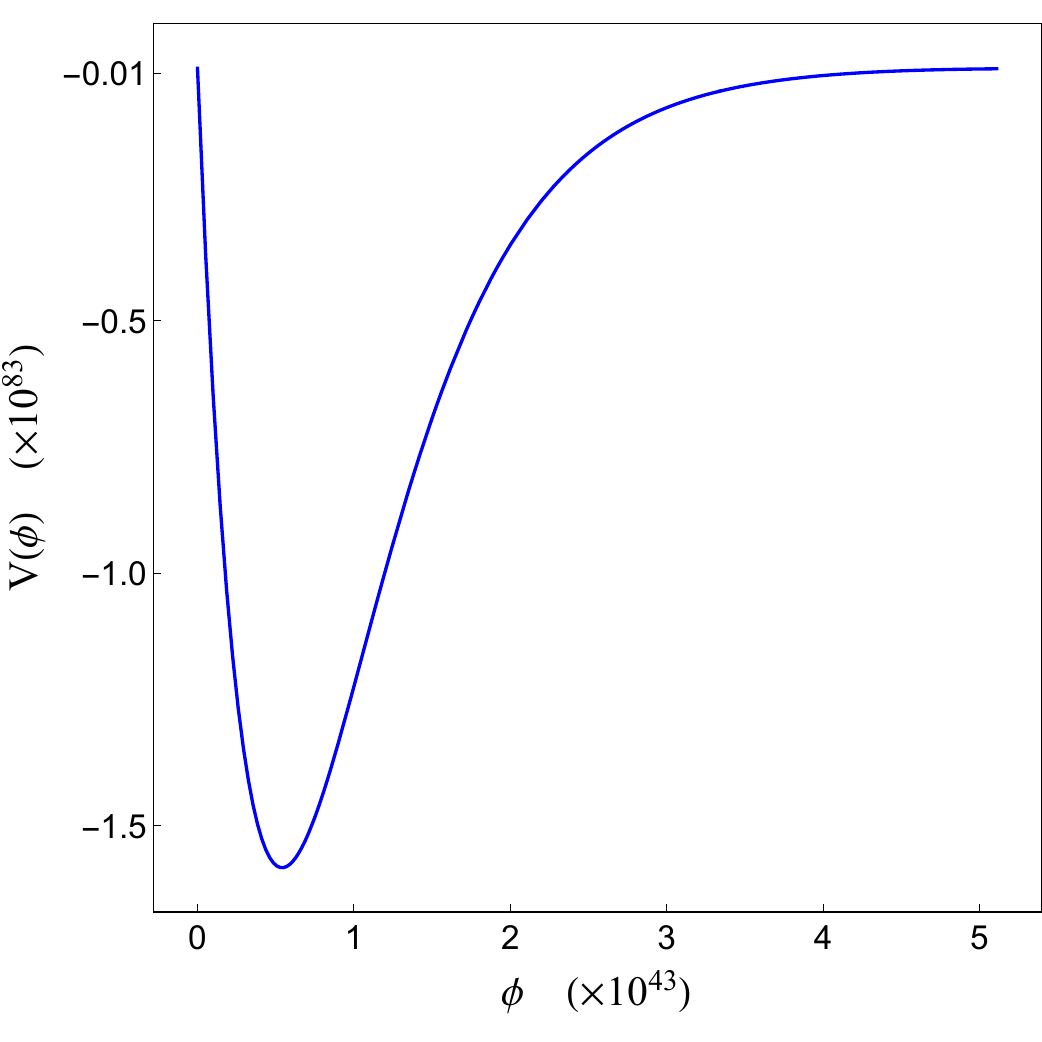}
		\caption{The parametric form from numerical solution of Eqs. (\ref{eq:1st}) and (\ref{eq:2nd}) for field potential for NFW density profile with Milky way galaxy parameters.}
		\label{fig:VphiNFWMilkyWay}
	\end{figure}

	For NFW profile, where the NFW profile parameters for Milky way are given in Table. (\ref{table:NFW-parameters}), we obtain the numerical solutions of Eqs. (\ref{eq:1st}) and (\ref{eq:2nd}). The obtained solutions are then plotted in Figs. (\ref{fig:Phi-V-NFWMilkyWay}) and (\ref{fig:VphiNFWMilkyWay})
	
	By looking at the graphs, we see that the field increases in the increasing direction of radial distance. The field potential on the other hand (as a function of radial distance as well as in parametric form) first decreases and then increases, giving a potential well. Therefore in this scenario, we see a concentration of DM scalar field in the well, and outside well a \emph{slow rolling field} towards the potential well is obtained.
	
	%-------------------------------
	%-------------------------------
	%-------------------------------
	\subsection{Numerical NFW M31 solution}\label{Numerical-NFW-M31-Solution}
\begin{figure}%[h!]
		\centering
		\begin{minipage}{0.48\linewidth}
			\centering
			{\includegraphics[height=0.8\textwidth]{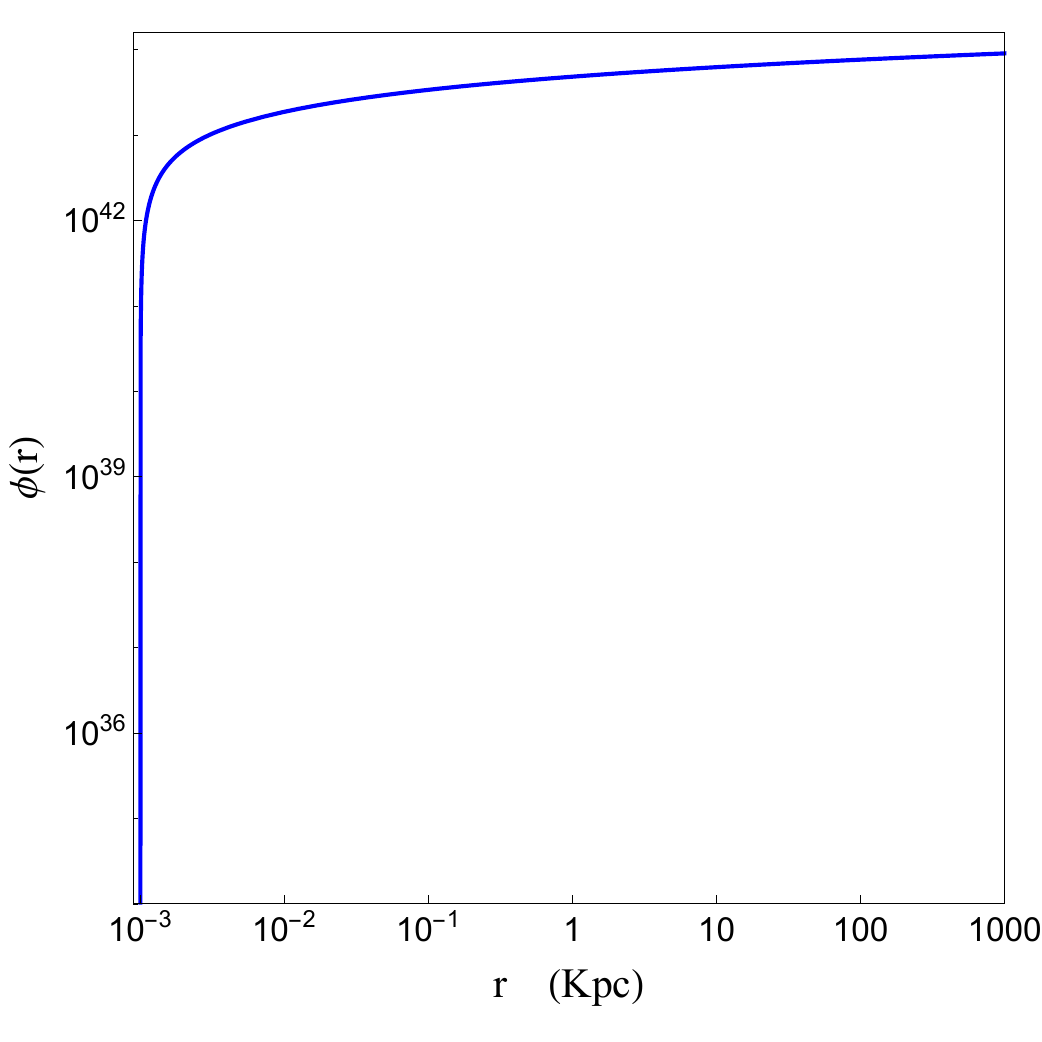}}
			\subcaption{The DM field.}
			\label{fig:phirNFWM31}
		\end{minipage}
		\hspace{0.1cm}
		\begin{minipage}{0.48\linewidth}
			\centering
			{\includegraphics[height=0.8\textwidth]{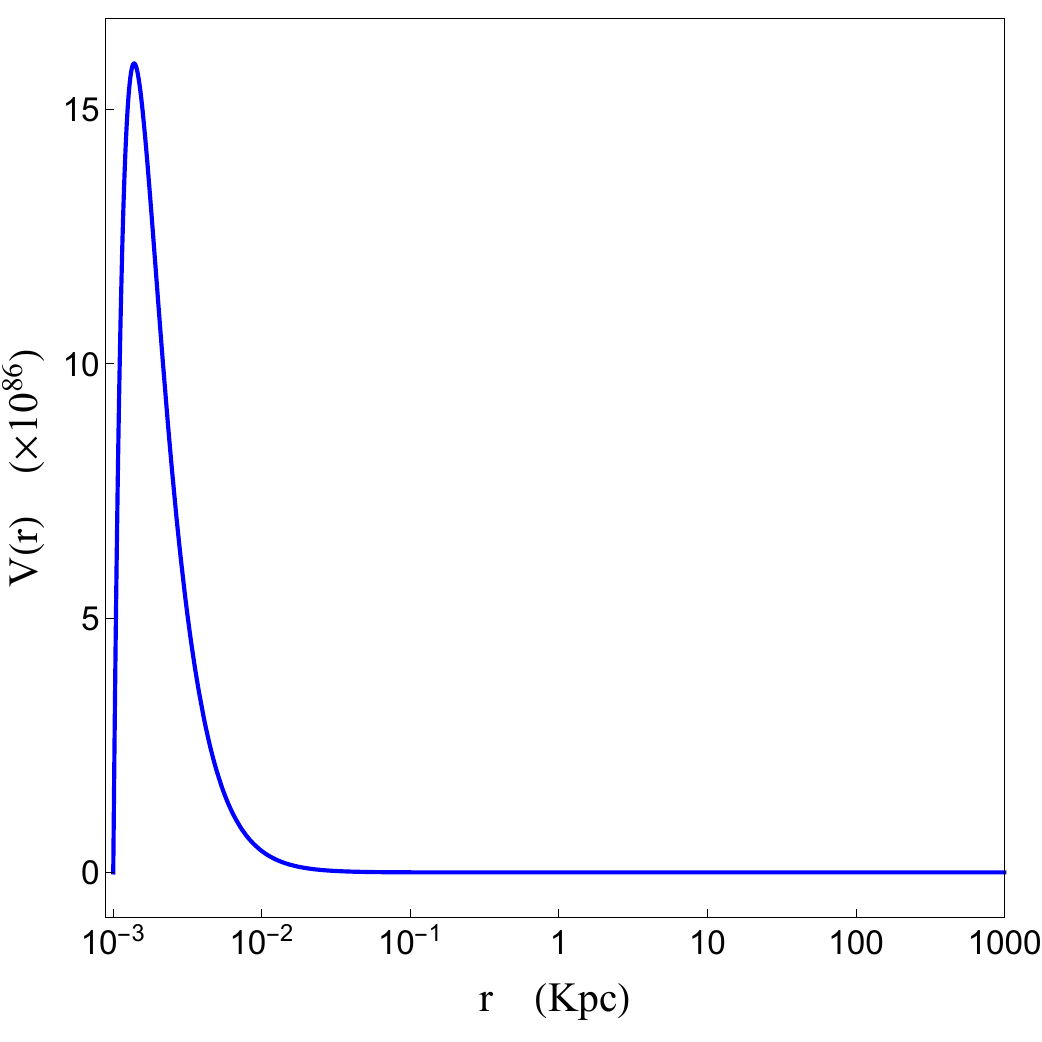}}
			\subcaption{The field potential.}
			\label{fig:VrNFWM31}
		\end{minipage}
		%\\[2ex]
		%	\begin{minipage}{0.95\linewidth}
		%		\centering
		%		\includegraphics[height=0.6\textwidth]{Plots/Scalar-DM/VphiNFWMilkyWay.pdf}
		%		\caption{The field potential as a function of field.}
		%		\label{fig:VphiNFWMilkyWay}
		%	\end{minipage}
		\caption{The scalar field properties for NFW density profile with M31 galaxy parameters.}
		\label{fig:Phi-V-NFWM31}
	\end{figure}

\begin{figure}%[h!]
		\centering
		\includegraphics[height=0.4\textwidth]{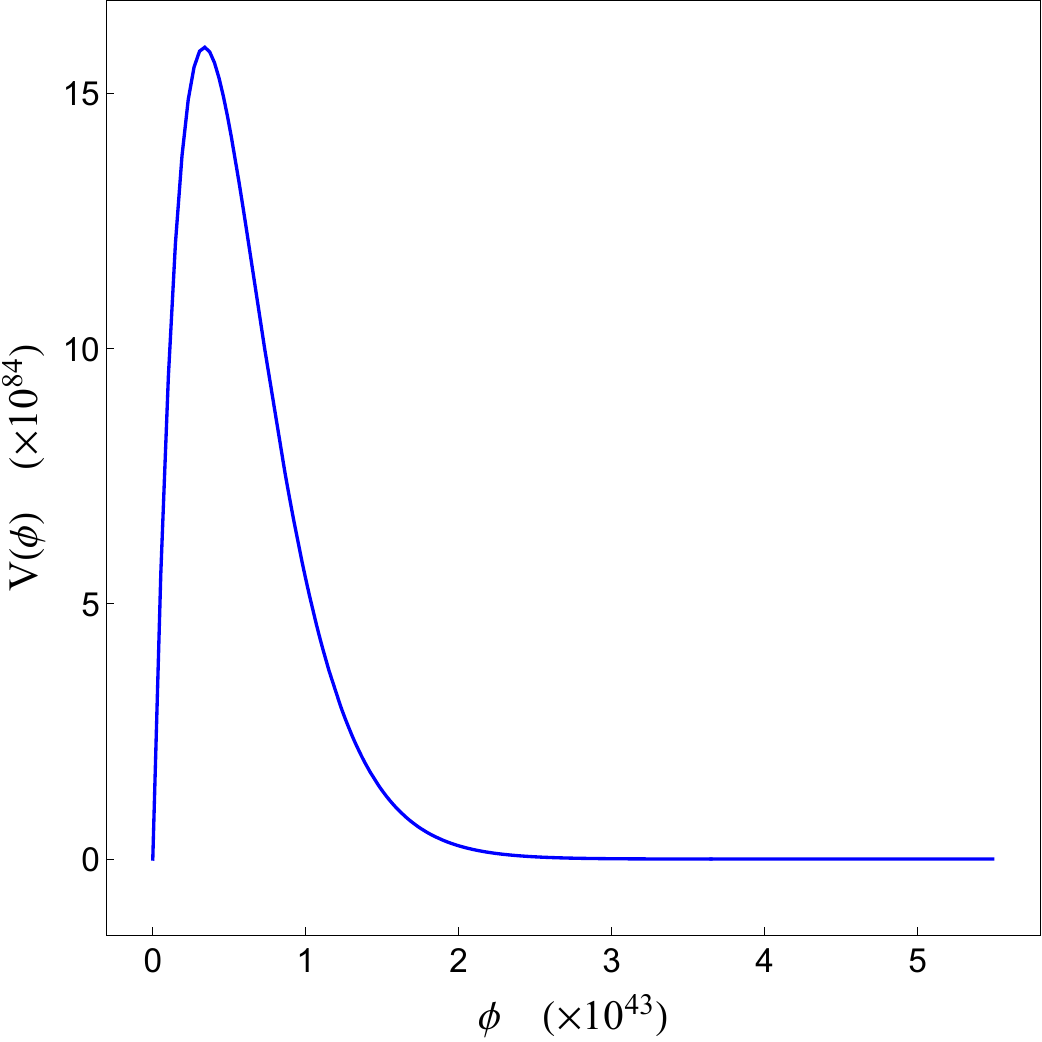}
		\caption{The parametric form from numerical solution of Eqs. (\ref{eq:1st}) and (\ref{eq:2nd}) for field potential for NFW density profile with M31 galaxy parameters.}
		\label{fig:VphiNFWM31}
	\end{figure}

	For NFW profile, where the NFW profile parameters for M31 are given in Table. (\ref{table:NFW-parameters}), we obtain the numerical solutions of Eqs. (\ref{eq:1st}) and (\ref{eq:2nd}). The obtained solutions are then plotted in Figs. (\ref{fig:Phi-V-NFWM31}) and (\ref{fig:VphiNFWM31})
	
	By looking at the graphs, we see that the field increases in the increasing direction of radial distance as in the previous NFW Milky Way scenario. The field potential on the other hand (as a function of radial distance as well as in the parametric form) first increases and then decreases, giving a potential barrier. This behavioral form of potential is exactly opposite to what we encountered in the NFW  Milky Way scenario. Therefore for M31, we expect a \emph{DM halo} at the position of the barrier. As M31 is a type SAb spiral galaxy in which spirals are significant. Close to and on the location of the spirals the normal matter is more concentrated thus DM is less concentrated which physically is the DM halo. Mathematically, we expect the DM potential peak at the location of the DM halo which would represent a less concentrated DM region. On the contrary, one explanation of the opposite behavior for the milky way galaxy could the the following: since the milky way is the barred spiral galaxy where the most of the normal mass is concentrated near the galactic center whereas a single arm extends(rounds) the galaxy with far less normal matter distribution, this allows for a much wider space for the DM to distribute which allows for a stable DM orbits and therefore a DM potential well.
	
	%%%%%%%%%%%%%%%%%%%%%%%%%%%%%%%%%%%%%%%%%%%%%%%%%%%%%%%%%%%%%%%%%%%%%%%%%%%%%%%%%%%
	%%%%%%%%%%%%%%%%%%%%%%%%%%%%%%%%%%%%%%%%%%%%%%%%%%%%%%%%%%%%%%%%%%%%%%%%%%%%%%%%%%%
	%\newpage
	%%%%%%%%%%%%%%%%%%%%%%%%%%%%%%%%%%%%%%%%%%%%%%%%%%%%%%%%%%%%%%%%%%%%%%%%%%%%%%%%%%%%%%%%
	\newpage
	\section{Conclusions and Discussion}\label{conclusions}
	
	In this article, we started of with a metric that describes the constant rotational velocity region in spherically symmetric and static space-time. After showing that obtaining analytic solution is not possible, we then obtained the numerical solution to the field equations using different scenarios which describe the galactic matter density profile to get an appropriate estimate of $f(\mathcal{R})$ gravity that could be an alternative to the so-called dark matter. The resultant graphs showed very slight modification to the Einstein's gravity for vacuum, and significant deviation from Einstein's gravity can be an alternative to the dark matter for power law density, simple model of a galaxy with a core and NFW profiles. For density profile which given constant tangential velocity at any distance from the galactic center (Eq. (\ref{eq:Power-law-density-profile})), a permanent deviation from Einstein's gravity is obtained meaning that we do not have $f(\mathcal{R})=\mathcal{R}$ at the galactic edge. As said in the last section, the correction up to $r\approx 200\text{Kpc}$ in vacuum we find is negative i.e. $f(\mathcal{R})<\mathcal{R}$ for $r>200\text{Kpc}$ the correction is positive i.e. $f(\mathcal{R})>\mathcal{R}$, shown in Table \ref{table:rRf}. %Although we are not giving any numerical result beyond $r=500\text{Kpc}$ yet it is worth mentioning that in our numerical solutions after $r\gtrapprox 885\text{Kpc}$, the modifications in the $f(\mathcal{R})=\mathcal{R}$ becomes positive i.e. for $r \gtrapprox 885\text{Kpc}$, $f(\mathcal{R})>\mathcal{R}$; where as this same shift is observed at $r=35.3\text{Kpc}$ for density profile given by Eq. (\ref{eq:Power-law-density-profile}). %In concluding suggests that the a very slight negative correction to the $f(\mathcal{R})=\mathcal{R}$ can explain the constant galactic rotation curves.
	
	It was also proven analytically in \citet{Usman2016} that in vacuum when $\lambda$ is a constant then, in the limit that $m^2$ and higher order terms in $m$ can be neglected, the solution of Eq. (\ref{eq:DMeq1}) and (\ref{eq:DMeq2}) is of the form of $\mathcal{R}^{1-\epsilon}$ where $\epsilon$ is a positive valued function of $m$, again implying a negative correction to the GR. %Thus our results here are consistent with the earlier results obtained in the limit when $\lambda$ is a constant.
	
	%It is well known that in the modified Newtonian dynamics (MOND), we only require acceleration of the order of $10^{-10}m/s^2$ to explain the observed rotational velocity curves \citet{1984AA136L21S,Milgrom:1983ca,PhysRevD.70.083509,PhysRevD.71.069901}. This small value of acceleration can be regarded as only a small deviation of galactic rotational curves from general relativity. Also, the Pioneer anomaly is of the order of $10^{-10}m/s^2$, which again implies the small modifications in the general relativity. %only the small deviation from $f(\mathcal{R})=\mathcal{R}$.
	
	%On the other hand, recent laboratory tests have confirmed that Newton?s second law is in good agreement with accelerations of the order of $5\times 10^{-14} m/s^{-2}$ [38]. Similar constraints have also been obtained for the inverse square law, where it was shown that Newton?s law holds down to a length scale of 56 lm [39]. Hence, it follows from these observations and experiments that in the $\mathcal{R}^n$ modified theories of gravity the parameter n should be of the form $1+\delta$ with $\delta\ll 1$.
	
	The weak field limit of the $f(\mathcal{R})$ gravity models has been discussed for star like objects in \citet{PhysRevD.75.124014,PhysRevD.77.108501,PhysRevD.77.108502}. If we assume that $f(\mathcal{R})$ is an analytical function at the constant curvature $\mathcal{R}_0$, that $m_\phi/r \ll 1$, where $m_\phi$ is the effective mass of the scalar degree of freedom of the theory, and that the fluid is pressureless, the post-Newtonian potentials $\Psi(r)$ and $\Phi(r)$ are obtained for a metric
	\begin{equation*}
	\text{d}s^2=-\left(1-2\Psi(r)\right)\text{d}t^2+\left(1+2\Phi(r)\right)\text{d}r^2+r^2\text{d}\Omega^2.
	\end{equation*}
	In that way, the behavior of $\Psi(r)$ and $\Phi(r)$ outside the star can be estimated. This then gives the value of the post-Newtonian parameter $\gamma$ to be $1/2$ \citet{Bohmer2008386}. From the solar system observations, we know that $\gamma=1$ \citet{Bohmer2008386}. This inconsistency between expected and measured value of $\gamma$ then is used to rule out most of the modified gravity models. T. Chiba et. al. %, T.L. Smith and A.L. Erickcek 
	have done the same analysis for the $f(\mathcal{R})$ gravity function of the form $f(\mathcal{R})=\left(\mathcal{R}/\alpha\right)^{1+\delta}$ in \citet{PhysRevD.75.124014}. They concluded that in the their section III. Case Studies ``...this analysis is incapable of determining whether $f(\mathcal{R})=\mathcal{R}^{1+\delta}$ gravity with $\delta\neq1$ conflicts with solar system tests.''. Thus, we can conclude that results presented %in Fig. (\ref{fig:fR}) and Table (\ref{table:rRf}) 
	does not contradict with the solar model tests.
	
	One should keep in mind that the graphs presented above are obtained in approximation that 
	\begin{enumerate}
		\item Vacuum, $\rho=0$.
		\item The rotational velocity of a particle around the galaxy in radial direction remains constant, Eq. (\ref{eq:Power-law-density-profile}).
		\item The rotational velocity of a particle around the galaxy first increases and then becomes constant in the increasing radial direction, Eq. (\ref{eq:Simple-model-for-a-galaxy-with-a-core}).
		\item Navarro, Frenk and White (NFW) profile, Eq. (\ref{eq:NFW-profile}).
	\end{enumerate}
	However, any of the above scenario is not exactly the case observed (but much close to the observations), trend of rotational velocity curve for different galaxies is only slightly different. The assumptions are valid for at least a significant region of the total tangential velocity profile. If one can establish an exact relation between the rotational velocity and distant from the galactic center then one can possibly obtain more accurate numerical results of the field equations. %In that case, we might even be able to find the exact solution to the field equations.
	
	The tangential velocity of the objects rotating around the galactic center is about $200-300\text{Km/s}$ \citet{Salucci11062007,Persic01071996,Borriello11052001} which give $m\approx (10^(-6))$. Changing the values slightly from this would not make any difference because of smallness of this value. Increasing the value significantly and we expect that the contribution from the normal matter part becoming less and less significant as evident from Eq. (\ref{eq:1st}) since its last term is inversely exponentially dependent on m. This dependence is natural as the tangential velocity would increase the DM dependence would increase significantly since the normal matter contribution is fixed and calculable from conventional observations methods. This always leads us to believe that increased tangential velocity can only be the result of the increased DM contribution in the galaxies.
	
	We also used the BD theory to make an analogy between the scalar DM theory and the modified gravity version. We found that in vacuum and for Milky way NFW profile the scalar DM potential has an attractor solution where the well lies close to the galactic center, this gives the similar increasing behavior of the potential as the Newtonian gravity. For M31 NFW profile, the scalar DM potential has an repulsive solution where the delta type function's peak lies close to the galactic center. Thus, for M31 th DM does not lie close to the galactic center where as for Milky Way it does. For power law matter density profile, uthe potential increases in the increasing direction of radial distance. The interesting result of simple galactic model tell that the small BSM DM coalesce around the galactic center in a form of galactic rings with increasing ring radius and decreasing density of the DM as the radius of the ring increases. This result indicate a completely different gravitational interaction of matter, which needs to be explored further.
	
	In concluding, we found that regions with flat galactic rotations curves do not require any kind of dark matter if we try to explain these in $f(\mathcal{R})$ gravity. The constant rotation curves are then a consequence of the additional geometrical structure provided by the modified gravity function $f(\mathcal{R})$. We found that a modification to the Einstein-Hilbert Lagrangian may account for the existence of ``dark matter'' (in some scenarios it is slight whereas in others it is significant). In the $f(\mathcal{R})$ gravity modified theory, in principle we can rewrite the field equations in terms of the Einstein tensor, $\mathcal{R}$, and interpret the remaining term(s) as a geometrical energy-momentum tensor which then gives us the option not to include the exotic type of matter called dark matter. In this way, the presence of the remaining terms can provide us with an elegant geometric interpretation of the dark matter.
	\emph{But} if th DM exists, the $f(\mathcal{R})$ gravity provides a very useful information about its properties and behavior in galactic environment.
	%%%%%%%%%%%%%%%%%%%%%%%%%%%%%%%%%%%%%%%%%%%%%%%%%%%%%%%%%%%%%%%%%%%%%%%%%%%%%%%%%%
	%\section{Acknowledgment}
	%This work is supported by {\textit{National University of Sciences and Technology (NUST), Sector H-12 Islamabad 44000, Pakistan}}.
	%%%%%%%%%%%%%%%%%%%%%%%%%%%%%%%%%%%%%%%%%%%%%%%%%%%%%%%%%%%%%%%%%%%%%%%%%%%%%%%%%%
	
	\nocite{*}

	\bibliography{MuhammadUsman.bib}% Produces the bibliography via BibTeX.
	
\end{document}